\documentclass[iop,dvipdfmx,apj,twocolappendix,numberedappendix]{emulateapj}
\usepackage[varg]{txfonts}
\usepackage{bm,ulem}

\newcommand{\pr}[1]{\ensuremath{\left(#1\right)} }
\newcommand{\Pf}[2]{\ensuremath{\left( \frac{#1}{#2} \right)} }
\newcommand{\avg}[1]{\ensuremath{\left\langle#1\right \rangle} }
\newcommand{\dfrac}[2]{ {\displaystyle\frac{#1}{#2}} }
\newcommand{\eqref}[1]{(\ref{#1})}

\defcitealias{Okuzumi2015The-Nonlinear-O}{OI15}
\newcommand{\myemail}{mori.s@geo.titech.ac.jp}

\begin{document}
\title{ Electron Heating in the Magnetorotational Instability: Implications for Turbulence Strength in Outer Regions of Protoplanetary Disks}
\shorttitle{Electron Heating in the MRI}
\author{Shoji Mori\altaffilmark{1,2} and Satoshi Okuzumi\altaffilmark{1,3}}
\shortauthors{Mori \& Okuzumi}
\affil{\altaffilmark{1}Department of Earth and Planetary Sciences, Tokyo Institute of Technology, Meguro-ku, Tokyo, 152-8551, Japan; \myemail}
\affil{\altaffilmark{2}Department of Physics, Nagoya University, Nagoya, Aichi 464-8602, Japan}
\affil{\altaffilmark{3}Jet Propulsion Laboratory, California Institute of Technology, Pasadena, CA 91109, USA}

\begin{abstract}
The magnetorotational instability (MRI) drives vigorous turbulence in a region of protoplanetary disks where the ionization fraction is sufficiently high. 
It has recently been shown that the electric field induced by the MRI can heat up electrons and thereby affect the ionization balance in the gas. In particular, 
in a disk where abundant dust grains are present, the electron heating causes a reduction of the electron abundance, thereby preventing further growth of the MRI. 
By using the nonlinear Ohm's law that takes into account electron heating, 
we investigate where in protoplanetary disks this negative feedback between the MRI and ionization chemistry becomes important. 
We find that the ``e-heating zone,'' the region where the electron heating limits the saturation of the MRI, 
extends out up to 80 AU in the minimum-mass solar nebula with abundant submicron-sized grains. 
This region is considerably larger than the conventional dead zone whose radial extent is $\sim20$ AU in the same disk model. 
 Scaling arguments show that the MRI turbulence in the e-heating zone should have a significantly lower saturation level. 
Submicron-sized grains in the e-heating zone are so negatively charged that their collisional growth is unlikely to occur.  
Our present model neglects ambipolar and Hall diffusion, but our estimate shows that ambipolar diffusion would also affect the MRI in the e-heating zone.
\end{abstract}


\section{Introduction}
Magnetorotational instability \citep[MRI;][]{Balbus1991A-powerful-loca} is widely regarded as a mechanism driving turbulence in protoplanetary disks. 
Vigorous MRI turbulence provides an effective viscosity that allows disk accretion 
at a rate consistent with observations \citep{Hawley1995Local-Three-dim,Flock2011Turbulence-and-}.
MRI also generates outflows from the disk surface \citep{Suzuki2009Disk-Winds-Driv,Suzuki2010Protoplanetary-,Bai2013Wind-driven-Acc,Lesur2013The-magnetorota,Fromang2013Local-outflows-}.
In addition, MRI turbulence have many important effects on the evolution of solid particles and planet formation. The effects include diffusion of small grains \citep{Carballido2005Diffusion-coeff},
concentration of larger solid particles \citep{Johansen2006Gravoturbulent-}, 
enhancement of particles' relative velocity that could lead to their collisional disruption \citep{Carballido2010Relative-veloci} of meter-sized bodies, 
and random migration of planetesimals  \citep[e.g.,][]{Laughlin2004Type-I-Planetar,Nelson2010On-the-dynamics}.

However, in weakly ionized protoplanetary disks, 
the saturation level of MRI turbulence depends strongly on non-ideal MHD effects and hence on 
the ionization state of the disks. 
Since thermal ionization is relevant only close to the central star \citep{Umebayashi1983The-Densities-o}, 
the dominant part of the disks is ionized only by high-energy sources such as 
stellar X-rays \citep{Glassgold1997X-Ray-Ionizatio} and galactic cosmic rays 
\citep{Umebayashi1981Fluxes-of-Energ}.
Deep inside the disks, the ionization fraction is significantly low 
because these ionizing radiations are attenuated and because recombination proceeds fast.

The low ionization fraction gives rise to fast Ohmic dissipation that stabilizes the MRI 
\citep{Sano1999Magnetorotation}.
Such a region is called the ``dead zone''  \citep{Gammie1996Layered-Accreti}.
The MRI is also suppressed by ambipolar diffusion near the surface of the disks \citep{Desch2004Linear-Analysis,Bai2011Effect-of-Ambip,Dzyurkevich2013Magnetized-Accr}.
The Hall effect can either stabilize or destabilize the MRI depending on the orientation of 
the magnetic field relative to the disk rotation axis \citep{Wardle1999The-Balbus-Hawl,Wardle2012Hall-diffusion-,Bai2014Hall-effect-Con}.

A number of studies have quantified how far the dead zone extends in protoplanetary disks. 
\citet{Gammie1996Layered-Accreti} assumed that the MRI is stable in a region where the column 
density exceeds the attenuation depth ($\approx 100~{\rm g~cm^{-2}}$) of galactic cosmic rays. 
More sophisticated models that incorporate ionization and recombination 
\citep[e.g.,][]{Sano2000Magnetorotation,Semenov2004Reduction-of-ch,Ilgner2006On-the-ionisati,Bai2009Heat-and-Dust-i,Dzyurkevich2013Magnetized-Accr}
showed 
that the MRI can be inactive even at lower column densities, with the predicted dead zone extending 
to $\sim 20~{\rm AU}$ from the central star when micron-sized dust grains are abundant in the disks. 
The abundance of small grains is relevant here because these particles efficiently sweep up plasma particles 
and thus lower the ionization fraction. 

All the previous studies on the dead zone assumed that vigorous MRI turbulence is sustained outside the zone. 
However, \citet[][henceforth \citetalias{Okuzumi2015The-Nonlinear-O}]{Okuzumi2015The-Nonlinear-O} suggested that the ionization fraction would be decreased by electric fields induced by MRI turbulence.
In a magnetorotationaly unstable region, the MRI turbulence generates strong electric fields associated with the growth of magnetic fields.
Plasma particles are accelerated by the strong electric fields and are scattered isotropically by collisions with neutral gas particles, leading to increase of their thermal velocity. 
In particular, electrons are more easily heated compared to ions because light particles are easily scattered.
Therefore, the sufficiently developed electric fields of MRI turbulence increase electron temperature in a weakly ionized gas \citep[{\it electron heating}; ][]{Inutsuka2005Self-sustained-}.
The heated electrons frequently collide with and stick to dust grains.
As a result, the electron heating decreases the ionization fraction.

Reduction of ionization fraction caused by the electron heating amplifies Ohmic dissipation, and, as a result, MRI turbulence may be suppressed.
This negative feedback that the MRI growth causes suppression of the MRI can be a saturating mechanism of MRI.
Although previous simulations \citep[e.g.,][]{Sano2002aThe-Effect-of-t,Simon2015Magnetically-dr,Flock2015Gaps-rings-and-,Bai2015Hall-Effect-Con} in a well ionized regions showed that MRI turbulence sustains a fully developed state, the turbulence strength may be suppressed at a lower saturation level by the effect of the electron heating.
However, it is not clear whether electric fields can sufficiently grow to heat up electrons before the MRI fully develops, and whether the decrease of the turbulent saturation level is meaningful.
The goal in this paper is to investigate where in protoplanetary disks the electron heating affects MRI turbulence and estimate how the saturation level would be suppressed.
This investigation is the first-step towards exploring the importance of the electron heating in protoplanetary disks.
In this study, we take into account only the Ohmic dissipation and neglect the other non-ideal effect of MHD for simplicity.

In Section \ref{sec:Method}, we present the disk model, simplified plasma heating model, and ionization balance.
In Section \ref{sec:Zones}, we present some conditions for MRI growth and some criteria for mapping of turbulent state in a disk.
We also briefly summarize the turbulent state and calculation steps. 
In Section \ref{sec:Results}, we show where the electron heating affects MRI turbulence. We also consider cases with various parameters. 
In Section \ref{sec:Suppress}, we estimate how the electron heating suppresses MRI turbulence.
In Section \ref{sec:CB}, we discuss the effect of heated electrons on the electric repulsion and the collisional growth of dust grains.
In Section \ref{sec:Discussion}, we discuss the effects neglected in our study.
In Section \ref{sec:Summary}, we present a summary of the results.

\section{Disk and Ionization Models} \label{sec:Method}

\subsection{Disk Model} \label{ssec:Disk-Model}
We consider a gas disk around a solar-mass star.
We assume that the surface density of the disk gas obeys a power law
\begin{equation}
\Sigma( r ) = 1.7 \times 10^{3} f_{\Sigma} \pr{\frac{r}{1~\mathrm{AU}}}^{-3/2} ~ \mathrm{g\, cm^{-2}},
\end{equation}
where $r$ is the distance from the central star, and $f_\Sigma$ is a dimensionless parameter.
The choice of $f_\Sigma = 1$ corresponds to the minimum-mass solar nebula (MMSN) model of \citet{Hayashi1981Structure-of-th}, which we take as the fiducial model. 

We assume that the disk is optically thin and give the temperature profile as \citep{Hayashi1981Structure-of-th}
\begin{equation}\label{eq:T}
T( r )=280\Pf{r}{ 1~\mathrm{AU} }^{- 1/2} ~\mathrm{K},
\end{equation}
where the central star is assumed to have the solar luminosity.  

The sound speed is given by
$c_{s}=\sqrt{kT/ m_{n}}$, where $ m_{n}$ is the mass of a neutral gas particle, and $k$ is the Boltzmann constant.
Assuming $m_n = 2.34~{\rm amu}$ and using Equation \eqref{eq:T}, we have 
\begin{equation}\label{eq:cs}
		c_{s}( r )=1.0\times 10^{5} \Pf{r}{1~ \mathrm{AU} }^{-1/4} ~\mathrm{cm\,s^{-1}}.
\end{equation}

We assume that the gas disk is hydrostatic in the vertical direction and give the vertical distribution of the gas density as  
\begin{equation}\label{eq:rho}
\rho( r, z )=\rho_{c}( r )\exp{\pr{- \frac{z^{2}}{2 H^{2}} }},
\end{equation}
where $\rho_{c}$ is the mid-plane density and 
$H \equiv c_s/\Omega$ is the gas scale height with 
$\Omega = 2.0\times 10^{-7}~(r/1~{\rm AU})^{-3/2}~{\rm s}^{-1}$ 
being the orbital frequency (note that a solar-mass star is assumed).  
Using the relation 
$\Sigma = \int^{\infty}_{- \infty} \rho dz = \sqrt{2 \pi}H\rho_{c}$, we have
\begin{equation}\label{eq:rhoc}
	 	 \rho_{c}( r )= 1.4\times 10^{-9} f_{\Sigma} \Pf{r}{1~\mathrm{AU} }^{-11/4}    ~\mathrm{g~cm^{-3}}.
\end{equation}
Thus, the number density of gas particles $n_{n} = \rho/m_{n}$ is given as
\begin{equation}\label{eq:}
	n_{n} (r, z) = 3.5\times 10^{14} f_{\Sigma} \times \Pf{r}{1  ~\mathrm{AU} }^{-11/4}   \exp{\pr{- \frac{z^{2}}{2 H^{2}} }}~ \mathrm{cm^{-3}}. 
\end{equation}

As we will describe in Section \ref{ssec:Criteria}, the criteria for MRI depends on 
the magnetic field strength in the disk. 
Following \citet{Sano2000Magnetorotation}, we consider a net (large-scale) vertical field $B_{z0}$
threading the disk and specify its strength with the plasma beta at the midplane, 
$\beta_c \equiv 8\pi\rho_c c_{s}^2/B_{z0}^2$. If we use Equations~\eqref{eq:cs} and \eqref{eq:rhoc}, 
the net vertical field strength can be expressed as 
\begin{equation}\label{eq:B}
	B_{z0}(r) = 0.59 f_{\Sigma}^{1/2} \Pf{\beta_{c}}{1000}^{-1/2} 
	\Pf{r}{1 ~\mathrm{AU} }^{-13/8}~ \mathrm{G}.
\end{equation}
For simplicity, we will assume that $\beta_c$ is constant in the radial direction.  

The charge reaction model adopted in this study takes into account the effects of grain charging on the ionization balance. 
For simplicity, we assume that dust grains are well mixed in the gas so that the dust-to-gas mass ratio $f_{dg}$ is a global constant. 
We also assume that the grains are spherical and single-sized with radius $a$ (taken as a free parameter) and internal density $\rho_\bullet$ (fixed to be $3~{\rm g~cm^{-3}}$). From these assumptions, the number density of dust grains $n_{d }$ is given by  $3f_{dg}\rho/(4\pi a^3 \rho_\bullet)$, which is expressed as  
\begin{eqnarray}\label{eq:}
	n_d (r, z) 
	&=& 1.1\times 10^{3} f_{\Sigma}  \Pf{f_{dg}}{0.01}  \Pf{\rho_{\bullet}}{3~  \mathrm{g~cm}^{-3}}^{-1}  \Pf{a}{0.1  ~\mathrm{\mu m}}^{-3} \nonumber\\
	&&\times \Pf{r}{1  ~\mathrm{AU} }^{-11/4}   \exp{\pr{- \frac{z^{2}}{2 H^{2}} }} ~ \mathrm{cm^{-3}}.
\end{eqnarray}

The disk is assumed to be ionized by galactic cosmic rays, stellar X-rays, and radionuclides.
The ionization rate can be expressed as
\begin{equation}\label{eq:}
	\zeta=\zeta_{\rm CR} +\zeta_{\rm XR} +\zeta_{\rm RN},
\end{equation}
where $\zeta_{\rm CR}$, $\zeta_{\rm XR}$, and $\zeta_{\rm RN}$ stand for the contributions from cosmic rays, X-rays, and radioactive decay, respectively.
The cosmic ray distribution is expressed as \citep{Umebayashi2009Effects-of-Radi}
\begin{eqnarray}\label{eq:}
	\zeta_{\rm CR} &=& \frac{ \zeta_{\rm CR,0} }{2} 
	 \left\{    \exp{    \pr{   -\frac{\chi              }{ \chi_{\rm CR} }    }      } \left[      1+\pr{   \frac{ \chi                }{ \chi_{\rm CR}}   }^{3/4}  \right] ^{-4/3}   \right.  \nonumber \\
	&&            \left. + \exp{    \pr{   -\frac{\Sigma - \chi}{\chi_{\rm CR} }    }      } \left[      1+\pr{   \frac{ \Sigma - \chi}{ \chi_{\rm CR}}   }^{3/4}    \right] ^{-4/3}   \right\} ,
\end{eqnarray}
where $\zeta_{\rm CR,0} = 1.0\times10^{-17}~{\rm s}^{-1}$ is the characteristic ionization rate of cosmic rays,
$\chi(r,z) = \int_z^\infty \rho(r,z') dz'$ is the vertical gas column density above height $z$, and
$\chi_{\rm CR} = 96~{\rm g~cm^{-2}} $ is the attenuation depth of ionizing cosmic rays. 
The ionization rate of X-rays is expressed as \citep{Bai2009Heat-and-Dust-i}
\begin{eqnarray}\label{eq:zetaxr}
\zeta_{\rm XR}&\! =\! &\frac{L_X}{10^{29}~ \rm\ erg\ s^{-1}}\left(\frac{r}{\rm 1~AU}\right)^{-2.2} \nonumber\\
&&\! \times \left\{   \zeta_{\rm XR, 1}  \left[    \exp{\pr{-\Pf{\chi}{\chi_{\rm XR,1}}^{0.4} }}      + \exp{\pr{ - \Pf{ \Sigma - \chi }{\chi_{\rm XR,1}}^{0.4}   }}   \right]    \right.    \nonumber\\
&&\!   \left.   +\zeta_{\rm XR, 2} \left[    \exp{\pr{-\Pf{\chi}{\chi_{\rm XR,2}}^{0.65} }}   +  \exp{\pr{ - \Pf{ \Sigma - \chi }{\chi_{\rm XR,2}}^{0.65} }}   \right]    \right\}\! , \hspace{4mm}
\end{eqnarray}
where $\chi_{\rm XR,1}$ and $\chi_{\rm XR,2}$ are taken to be $6 \times 10^{-3}{\rm ~g~cm^{-2}}$ and $3 {\rm~g~cm^{-2}}$ respectively, $\zeta_{\rm XR, 1}$ and $\zeta_{\rm XR, 2}$ are taken to be $6\times 10^{-12}  ~\mathrm{s}^{-1} $ and $1\times 10^{-15}~\mathrm{s}^{-1}$ respectively.
We take $L_x = 2\times 10^{30}~{\rm erg~s^{-1}}$ in accordance with the median X-ray luminosity of solar-mass young stars \citep{Wolk2005Stellar-Activit}.
The ionization rate of the radionuclide is expressed as \citep{Umebayashi2009Effects-of-Radi}
\begin{equation}\label{eq:}
\zeta_{\rm RN}=7.6\times 10^{-19} \Pf{f_{dg}}{0.01} ~ \mathrm{s^{-1}}.
\end{equation}

\subsection{Simplified Plasma Heating Model}
As we will describe in Section \ref{ssec:Criteria}, the criterion for MRI depends on 
the ionization fraction in the disk. 
We employ a simple ionization model proposed by \citetalias{Okuzumi2015The-Nonlinear-O} 
to calculate the ionization fraction taking into account plasma heating by a strong electric field. 
The model determines the ionization fraction of the gas at each location of a disk 
from the balance between ionization by external high-energy sources (e.g., cosmic rays and X-rays),
recombination in the gas phase, and adsorption of ionized gas particles onto dust grains.  
The rates of recombination and adsorption generally depend on the temperatures of ions and electrons, 
$T_i$ and $T_e$.
Previous ionization models assumed that $T_i$ and $T_e$ are equal to the neutral gas temperature $T$. 
By contrast, the model of \citetalias{Okuzumi2015The-Nonlinear-O} determines $T_i$ and $T_e$ 
as a function of the electric field strength $E$. 
For simplicity, positive ions are represented by the single species HCO$^{+}$,
which is good as a first-order approximation when heavy molecular ions 
that recombine through dissociation reactions dominate \citep{Umebayashi1990Magnetic-flux-l,Dzyurkevich2013Magnetized-Accr}.
We do not consider negative ions.
Although production of negative ions is rare in cool protoplanetary disks, 
electrons heated to $\ga 3~{\rm eV}$ can produce negative hydrogen ions H$^{-}$ via
dissociative electron attachment $ \mathrm{H} _{2} +  \mathrm{e}^{-} \to\mathrm{H} ^{-} +  \mathrm{H} $
 \citep{Wadehra1984Dissociative-at}.
However, H$^{-}$ would be instantly destroyed by CO, the most abundant molecule after H$_{2}$, 
via the reaction $ \mathrm{H} ^{-} +  \mathrm{CO}  \to  \mathrm{HCO} +  \mathrm{e} ^{-}$ \citep{Ferguson1973Rate-Constants-}.
For this reason, we may safely neglect the dissociative electron attachment during electron heating.

In this study, we make two further simplifications to the original model of \citetalias{Okuzumi2015The-Nonlinear-O}. 
Firstly, we calculate the electron temperature $T_e$ by solving the equations 
of momentum and energy conservation rather than 
by using the solution to the full Boltzmann equation.   
The rate coefficients for gas-phase recombination and plasma adsorption onto grains 
are then evaluated by approximating the velocity distribution function 
with a Maxwellian with temperature $T_e$. 
The approach greatly simplifies the analytic expressions of the rate coefficients 
that otherwise involve confluent hypergeometric functions (see Section 3 of~\citetalias{Okuzumi2015The-Nonlinear-O}). Such an approach was originally proposed by \citet{Hershey1939A-Theory-for-th} for calculating 
the mobility of heavy ions at a high electric field, and \citetalias{Okuzumi2015The-Nonlinear-O} 
followed this approach to compute the ion temperature $T_i$. 
In this study, we apply this approach to both $T_i$ and $T_e$. 
Secondly, we neglect the impact ionization of neutral molecules by electrically heated electrons
by assuming that the electron energy in MRI turbulence 
is well below the ionization potential of the neutrals ($\sim 10~{\rm eV}$).
The results of our calculations show that this assumption holds in most parts of 
protoplanetary disks. 

We denote the mean drift velocity and mean kinetic energy 
of a charged species $\alpha$ (= $i$ for ions, $e$ for electrons)
by $\avg{\bm{v}_{\alpha}}$ and $\avg{\epsilon_{\alpha}}$, respectively. 
In a weakly ionized gas with an applied electric field ${\bm E}$, 
the momentum and energy of the charged species are determined by 
the balance between the neutral gas drag and acceleration by the electric field \citep{Hershey1939A-Theory-for-th}. 
Explicitly, the solution of the momentum and energy balance equations  can be written as   
(Equations~(A9) and (A10) of \citetalias{Okuzumi2015The-Nonlinear-O})
\begin{equation}
\avg{{\bm v}_{\alpha}} = \dfrac{m_\alpha+m_n}{m_\alpha m_n} q_\alpha {\bm E} \Delta t_{\alpha} ,\label{eq:v-dri}
\end{equation}
\begin{equation}
\avg{\epsilon _\alpha} = \dfrac{3}{2}kT + \dfrac{(m_\alpha+m_n)^3}{2(m_\alpha m_n)^2} (q_\alpha E \Delta t_{\alpha})^2,\label{eq:eng}
\end{equation} 
where $q_{\alpha}$, $m_{\alpha}$, and $\Delta t_{\alpha}$ are the charge, mass and mean free time of the 
plasma particles (e.g., $q_{e} = -e$ and $q_{i} = e$, where $e$ is the elementary charge).
Since the magnetic field is neglected in this study, the mean drift velocity is parallel to the electric field. 
In a weakly ionized gas, the plasma mean free time is determined by neutrals gas particles,
\begin{equation}\label{eq:}
	\Delta t_{\alpha} = (n_{n}  \avg{\sigma_{\alpha n} v_{\alpha n}})^{-1},
\end{equation}
where $v_{\alpha n}$ is the relative velocity between a plasma particle and a neutral particle, and $\sigma_{\alpha n}$ is the momentum-transfer cross section for the plasma--neutral collision.
For electrons, $\sigma_{en}$ is approximately constant at low energies \citep{Yoon2008Cross-Sections-}, 
and therefore we may approximate $\avg{\sigma_{e n} v_{e n}}$ as $\sigma_{e n} \avg{v_{e n}}$. 
For ions, $\avg{\sigma_{i n} v_{i n}}$ is approximately constant owing to the polarization force 
between ions and neutrals \citep{Wannier1953Motion-of-Gaseo}.
Equations~\eqref{eq:v-dri} and \eqref{eq:eng} are exact only when $\Delta t_\alpha$ is constant, 
but still hold in a good accuracy even when $\Delta t_\alpha$ is velocity-dependent \citep{Wannier1953Motion-of-Gaseo}. 

The plasma temperature $T_\alpha$ is defined so that $3kT_{\alpha}/2$ 
is equal to the kinetic energy of random motion, 
$ \avg{\epsilon_{\alpha}} - m_{\alpha}\avg{\bm{v}_{\alpha}}^{2}/2$.
Using Equations (\ref{eq:v-dri}) and (\ref{eq:eng}), $T_\alpha$ can be written as  
\begin{equation}\label{eq:Talpha}
	T_{\alpha} = T + \dfrac{(m_\alpha+m_n)^2}{3 k m_\alpha^{2} m_n} (q_\alpha E \Delta t_{\alpha})^2.
\end{equation}
For electrons, we approximate $\avg{v_{en}}$ in $\Delta t_e$ with $\avg{v_e^2}^{1/2} = \sqrt{3kT_{e}/m_{e} }$.
This allows us to solve Equation~\eqref{eq:Talpha} with respect to $T_e$, and we obtain 
\begin{equation}\label{eq:Te}
	T_{e} = T  \pr{\frac{1}{2}+\sqrt{ \frac{1}{4} + \frac{2}{3}  \Pf{E}{E_{\rm crit}}^{2} }}, 
\end{equation}
where 
\begin{equation}\label{eq:Ecrit}
	E_{\rm crit} \equiv \sqrt{\frac{6 m_{e}}{m_{n}}}\frac{kT n_{n} \sigma_{en} }{e }
\end{equation}
is the critical field strength above which electron heating becomes significant. 
We have assumed $m_e \ll m_n$ in deriving Equation~\eqref{eq:Te}. 
For ions, Equation~\eqref{eq:Talpha}  directly gives
\begin{eqnarray}\label{eq:Ti}
    T_{i} &=& T \pr{ 1  +\frac{2 (m_{i}+m_{n})^{2} m_{e} }{m_{i}^{2} m_{n}^{2} } \frac{\sigma_{en}^{2} kT}{ \avg{\sigma_{i n} v_{i n}}^{2}} \Pf{E}{E_{\rm crit}}^{2}}, \nonumber\\
    &=& T \pr{ 1  + 7.6\times 10^{-7} \Pf{T}{100K} \Pf{E}{E_{\rm crit}}^{2}},
\end{eqnarray}
where we have set $\avg{\sigma_{i n} v_{i n}} = 1.6 \times 10^{-9}~{\rm cm}^{3}~{\rm s}^{-1}$ \citep{Nakano1986Dissipation-of-}
and $\sigma_{en} = 10^{-15}~{\rm cm}^{2}$ \citep{Yoon2008Cross-Sections-} in the second expression, and used $m_{i} = 29~{\rm amu}$.

\subsection{Ionization Balance and Accuracy of Simplified Approach}\label{ssec:ionz-bal}
We calculate the plasma densities in a protoplanetary disk 
taking into account grain charging. 
The equations that describe the ionization balance in a dusty disk are 
(Equations (32), (33) and (35) of \citetalias{Okuzumi2015The-Nonlinear-O})
\begin{equation}
\zeta n_{n} -K_{\mathrm{rec}}(T_{e}) n_{i}n_{e} -K_{de}(\phi, T_{e})n_{d}n_{e} = 0,\label{eq:ch-bal-e}
\end{equation}
\begin{equation}
\zeta n_{n} -K_{\mathrm{rec}}(T_{e}) n_{i}n_{e} -K_{di}(\phi, T_{i})n_{d}n_{i} = 0, \label{eq:ch-bal-i}
\end{equation}
\begin{equation}
n_{i}- n_{e} + Zn_{d} = 0,  \label{eq:ch-bal-Z}
\end{equation}
where $n_{e}$ and $n_{i}$ are, respectively, the number density of electrons and positive ions;
 $K_{\rm rec}$ is the gas-phase recombination rate; $K_{de}$ and $K_{di}$ are the adsorption rates of electrons and ions onto grains; $Z$ is the grain charge number; and $\phi$ is the coulomb potential on grain surface.
 $\phi$ is related to $Z$ as 
 \begin{equation}\label{eq:}
 		\phi = \frac{eZ}{a}.
 \end{equation}
As the collisional frequency, $K_{\rm rec}$ and $K_{de}$ depend on the electron temperature $T_{e}$, while $K_{di}$ depends on the ion temperature $T_{i}$.
$K_{de}$ and $K_{di}$ also depend on the coulomb potential of a grain surface $\phi$. 
 For HCO$^{+}$, the recombination rate $K_{\rm rec}$ is given by \citep{Ganguli1988Electron-temper}
\begin{equation}\label{eq:}
 	K_{\rm rec}(T_{e}) = 2.4 \times 10^{-7} \Pf{T_{e}}{300 ~\mathrm{K} }^{-0.69}
	~\mathrm{cm}^{3} ~\mathrm{s} ^{-1}.
\end{equation}
Approximating the ion velocity distribution by a Maxwellian with 
mean velocity $\avg{{\bm v}_{i}}$ and temperature $T_i$, $K_{di}$ is given by 
\citep[][\citetalias{Okuzumi2015The-Nonlinear-O}]{Shukla2002Introduction-to}
\begin{eqnarray}\label{eq:Kdi}
	K_{di} (\phi,T_{i}) &=& \pi a^2 \Biggl[ \sqrt{\frac{2 k T_i}{\pi m_i}}
\exp\left(-\frac{m_i\avg{{\bm v}_{i}}^2}{2 k T_i} \right)  
\nonumber \\
& &+ |\avg{{\bm v}_{i}}| 
\left( 1+\frac{k T_i + 2e|\phi| }{m_i \avg{{\bm v}_{i}}^2} \right) {\rm erf}
\left( \frac{|\avg{{\bm v}_{i}}| }{\sqrt{2 k T_i/m_{i}}}\right) \Biggr]. \hspace{5mm}
\end{eqnarray}
In this study, we also approximate the electron velocity distribution by a Maxwellian with temperature $T_e$. 
The drift velocity $\avg{{\bm v}_{e}}$ can be neglected here since the drift speed 
$|\avg{{\bm v}_{e}}|$ is generally much smaller than the random speed $\sim \sqrt{kT_e/m_e}$ 
owing to the smallness of $m_e/m_n$ \citep[see ][]{Golant1980Fundamentals-of,Lifshitz1981Physical-kineti}.
The electron adsorption rate coefficient $K_{de}$ is given by the simple expression 
\citep{Shukla2002Introduction-to}
\begin{equation}\label{eq:Kde}
	K_{de} (\phi,T_{e}) = \pi a^{2} \sqrt{\dfrac{8kT_{e}}{\pi m_{e}}}\times
	\left\{ \begin{array}{ll}
       \left(1 +\dfrac{e\phi}{kT_e} \right),  & \phi > 0, \\[3mm]
	\exp \left( \dfrac{e\phi}{kT_e} \right), &  \phi < 0.
\end{array} \right. 
\end{equation}
It should be noted that Equations (\ref{eq:Kde}) and (\ref{eq:Kdi}) assume perfect sticking of 
ions and electrons onto grain surfaces. This is a good approximation as long as the plasma temperatures 
are well below 100~{\rm eV} (see Section 3.2.2 of \citetalias{Okuzumi2015The-Nonlinear-O} for more discussion).

Equations~\eqref{eq:ch-bal-e}--\eqref{eq:ch-bal-Z} determine $n_e$, $n_i$ and $Z$ 
at each location in a disk as a function of $E$. 
We solve these equations using the procedure presented by \citet[][their Section~2.2; see also Section~3.2.4 of \citetalias{Okuzumi2015The-Nonlinear-O}]{Okuzumi2009Electric-Chargi}.

\begin{figure}
\centering
\includegraphics[width = 0.9\hsize,clip]{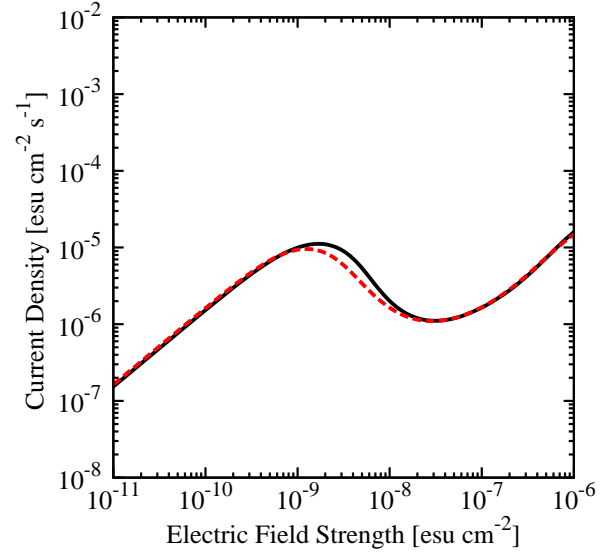}
\vspace{3mm}
\caption{
Test of the simplified plasma heating model presented in Section \ref{ssec:ionz-bal}. 
The solid curve shows the $J$-$E$ relation for 
`model C' of \citetalias{Okuzumi2015The-Nonlinear-O} 
derived using the exact electron velocity distribution 
(see Figure 10 of \citetalias{Okuzumi2015The-Nonlinear-O}), while
the dashed curve shows our reproduction based on the simplified approach. 
}
  \label{fig:hikaku}
\end{figure}

To test the accuracy of our simplified approach, 
we reproduce the current--field relation including plasma heating (the nonlinear Ohm's law of \citetalias{Okuzumi2015The-Nonlinear-O})
with adopting the calculation steps in \citetalias{Okuzumi2015The-Nonlinear-O}.
Current density is generally given by
\begin{equation}\label{eq:jdef}
	J (E) = q_{e} n_{e} \avg{ \bm{v} _{e}} + q_{i} n_{i} \avg{ \bm{v} _{i}}.
\end{equation}
Including plasma heating, the number densities depend on the electric fields strength $E$.
To obtain the current density, we first calculate plasma temperatures $T_{e}$ and $T_{i}$ from Equations (\ref{eq:Te}) and (\ref{eq:Ti}) in an applied electric field $E$.
We then calculate the number densities of plasma $n_{e}$ and $n_{i}$ from the ionization balance (Equation \eqref{eq:ch-bal-Z}).
We finally obtain the current density using Equations \eqref{eq:v-dri} and \eqref{eq:jdef}.
In Figure \ref{fig:hikaku}, we compare our result with the result of \citetalias{Okuzumi2015The-Nonlinear-O} for the parameter set `model C' of \citetalias{Okuzumi2015The-Nonlinear-O}.
We find that our calculation reasonably reproduces the previous result 
even at high field strengths  ($E \ga 10^{-9}~{\rm esu~cm^{-2}}$) where electron heating is significant.
The maximum relative difference between the two results is 37\%.

\section{Active, Dead, and E-heating Zone}\label{sec:Zones}

\subsection{Conditions for MRI Growth} \label{ssec:Criteria}
In the limit of ideal MHD, the criterion for the MRI is given by 
\citep{Balbus1991A-powerful-loca} 
\begin{equation}\label{eq:con-ide}
	\lambda _{\rm ideal} < H,
\end{equation}
where 
\begin{equation}\label{eq:lambda_ide}
\lambda _{\rm ideal} \equiv 2 \pi \frac{v_{Az}}{\Omega}
\end{equation} 
is the characteristic wavelength of the most unstable axisymmetric MRI modes, and 
$v_{Az} = B_z/\sqrt{4\pi\rho}$ and $B_z$ are the vertical components of the Alfven velocity
and magnetic field, respectively.
Equation~\eqref{eq:con-ide} expresses that the MRI operates when 
the lengthscale of the MRI modes is smaller than the vertical extent of the disk. 
When viewed as a function of $z$, $\lambda _{\rm ideal}$ increases with $z$ 
because $\rho$ decreases toward the disk surface. 
If we use Equation~\eqref{eq:rho}, the above MRI criterion can be rewritten 
in terms of height as 
\begin{equation}\label{eq:con-ide-z}
			z < \sqrt{2\ln{(\beta_{c}/8\pi^{2})} }~H \equiv H_{\rm ideal},
\end{equation}
where the height $H_{\rm ideal}$ defines the upper boundary of the MRI active zone. 

Inclusion of a finite Ohmic resistivity $\eta$ introduces another criterion for MRI growth. 
The criterion can be expressed in terms of the Elsasser number \citep{Turner2007Turbulent-Mixin}  
\begin{equation}\label{eq:Lambda}
	\Lambda \equiv \frac{v_{Az}^{2}}{\eta \Omega}.
\end{equation}
The instability grows when 
\begin{equation}\label{eq:Lam-con}
\Lambda > 1	
\end{equation}
 and decays when $\Lambda < 1$ \citep[e.g.,][]{Sano1999Magnetorotation}.

\subsection{Zoning Criteria}\label{ssec:DetTS}
Here we describe how to determine turbulent state at a position in protoplanetary disks.
Electron heating affects on the MRI turbulence when the ionization fraction is sufficiently decreased.
We express the condition that the heating takes place and affect MRI turbulence, 
and then summarize three turbulent states of MRI and steps of zoning a disk into the state.

For electron heating to take place, the field must be sufficiently amplified before MRI turbulence reaches a fully developed state that means the stop of MRI growth.
\citet{Muranushi2012Interdependence} performed a local unstratified resistive MHD simulation and found that the fully developed current density is 
\begin{equation}\label{eq:Jmax}
		J_{\mathrm{max}} = f_{\rm sat} \sqrt{\frac{\rho}{2\pi}}c \Omega,
\end{equation}
where $f_{\rm sat} \approx 10$ according to the results by \citet{Muranushi2012Interdependence}.
Here, we assume $f_{\rm sat}$ to be $f_{\rm sat}=10$ and the maximum current density is $J_{ \mathrm{max} }$.
Thus, when the current density reaches $J_{ \mathrm{max} }$ before electric field reaches the criterion for electron heating $E_{ \mathrm{crit} }$, 
MRI turbulence does not cause the electron heating.

As we will describe later in this section, we use current density to decide whether electron heating take place or not.
Therefore, we transform the condition for suppressing MRI into a form using current density.
We adopt $\Lambda = 1$ (Equation (\ref{eq:Lam-con})) as the criterion for suppressing MRI which is triggered by electron heating.
Using the electric conductivity $\sigma_{c}$ and the relation $\eta = c^2 /4 \pi \sigma_c$, the condition for sustaining MRI turbulence $\Lambda \gtrsim 1$ leads to a condition $\sigma_{c} \gtrsim c^{2} \Omega / (4 \pi v_{Az}^{2})$.
Under the Ohm's law $J(E) = \sigma_{c} E$, the condition can be rewritten as a lower limit to the current density 
\begin{equation}\label{eq:OC-dis-ex}
				J(E) \gtrsim  J_{\Lambda=1}(E),
\end{equation}
where 
\begin{equation}\label{eq:Jlam}
	J_{\Lambda=1}(E) \equiv \sigma_{c}(\Lambda = 1)E =  \frac{c^{2} \Omega }{4\pi v_{Az}^{2}}	 E.
\end{equation}

Using the above criteria, 
we can classify a region in protoplanetary disks into three different zones corresponding to three turbulent state of MRI.
\begin{enumerate}


	\item {\it Dead zone}.
	Because of the low ionization fraction, Ohmic dissipation suppress all the unstable MRI mode. 
	Suppressed MRI does not generate turbulence and also current density.
	We will refer to the region where MRI is completely suppressed as the ``dead zone''.
	In this case, the condition of Ohmic dissipation (Equation \eqref{eq:Jlam}) is satisfied with no MRI turbulence.

	\item {\it E-heating zone}.
Electric fields of MRI turbulence become sufficiently high for electron heating to be caused.
The Ohmic dissipation is amplified by the electron heating after the MRI grows.
We will refer to the region where electron heating affects MRI turbulence as the ``e-heating zone'', 
where the ``e'' refers to both ``{\it e}lectric field'' and ``{\it e}lectron.''
In this case, current density falls down the critical current density of Ohmic dissipation (Equation \eqref{eq:Jlam}).

	\item {\it Active zone}.
	MRI sustains fully developed turbulent state because the gas is sufficiently ionized so that Ohmic dissipation is not efficient. 
	We will refer to the region where vigorous MRI turbulence is sustained as the ``active zone'' in this study.
	In this case,	the current density $J$ reaches and sustains its maximum value $J_{\rm max}$ before electron heating reduces the MRI turbulence.

\end{enumerate}

\begin{figure}
\includegraphics[width = \hsize,clip]{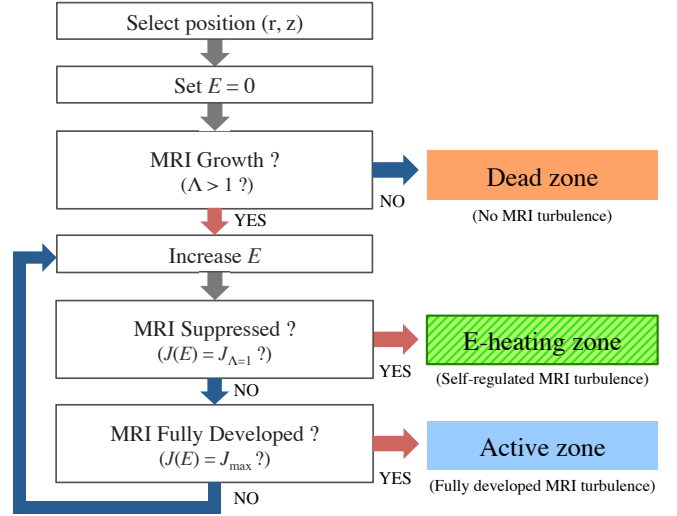}
  \caption{
Flow chart showing key steps of zoning a protoplanetary disk into the dead, active, and e-heating zones.
  }
  \label{fig:flow}
\end{figure}

We summarize the calculation steps for zoning the disk region under some assumptions. 
We assume that the electric field strength correspond to the activity of MRI turbulence since developed MRI generates strong electric fields. 
The growth of MRI implies increasing electric fields, and the decay of MRI implies decreasing electric fields. 
Furthermore, we also assume that magnetic fields are not varied by the MRI growth for simplicity.
Under these assumptions, we determine the turbulent state at the position with following steps (see Figure \ref{fig:flow}):
First, we select a calculated position in the region satisfying Equations (\ref{eq:con-ide-z}) of a disk.
We then calculate values at the position with setting $E = 0$.
When MRI is initially suppressed by Ohmic dissipation, i.e., $\Lambda < 1$ at $E = 0$, the positions belong to the dead zone.
During satisfying unstable condition, i.e., $\Lambda >1$, the electric field strength $E$ is increased from $E = 0$ with iterating until the turbulent state at the position is determined. 
We calculate current density $J(E)$ and assess some conditions in $E$.
When MRI turbulence causes electron heating and Ohmic dissipation become efficient, i.e., $J(E) = J_{\rm \Lambda = 1}$, the position belongs to the e-heating zone.
When MRI turbulence is fully developed, i.e., $J(E) = J_{\rm max}$, the position belongs to the active zone.
We conduct the above steps in the whole region in a disk, and zone a protoplanetary disk into the dead, active, and e-heating zones.

\section{Location of the E-heating Zone}\label{sec:Results}
We here predict the location of the e-heating zone in protoplanetary disks 
using the methodology described in Section~\ref{ssec:DetTS}. 
We conduct a parameter study varying the midplane plasma beta $\beta_c$,
grain size $a$, dust-to-gas mass ratio $f_{dg}$, and surface density scaling factor $f_\Sigma$.
Following \citet{Sano2000Magnetorotation}, we select the MMSN ($f_\Sigma = 1$ and $q=3/2$) 
with $a = 0.1~\micron$, $f_{dg} = 0.01$, and  $\beta_c = 1000$ as the fiducial model. 
We start out with this fiducial model in Section~\ref{ssec:fid}, and discuss the dependence 
on the parameters in the subsequent subsections. 
A summary of the parameter study is given in Table~\ref{tbl:beta}.
We also describe ion heating in Section \ref{ssec:IH}.

\begin{table}
\begin{center}
\caption{Sizes of the Dead and E-heating Zones for Various Parameter Sets}
\label{tbl:beta}\smallskip
\begin{tabular}{cccccc}
\hline \hline
$\beta_{c}$ & $a$ ($\rm \mu m$)& $f_{dg}$ & $f_{\Sigma}$  & \multicolumn{2}{c}{Outer radius (AU)}\\
& & & &  Dead zone &  E-heating zone \\
\hline
$10^{2}$ & $0.1$ & $10^{-2}$ & 1 &  18 & 74 \\
$10^{3}$ & $0.1$ & $10^{-2}$ & 1 &  24 & 82 \\
$10^{4}$ & $0.1$ & $10^{-2}$ & 1 &  34 & 82 \\
$10^{5}$ & $0.1$ & $10^{-2}$ & 1 &  56 & 82 \\ \hline
$10^{3}$ & $0.1$ & $10^{-2}$ & 1 &  24 & 82 \\
$10^{3}$ & $1$   & $10^{-2}$ & 1 &  11 & 39 \\
$10^{3}$ & $10$  & $10^{-2}$ & 1 &   8 & 19 \\
$10^{3}$ & $100$ & $10^{-2}$ & 1 &   8 & 11 \\ \hline
$10^{3}$ & $0.1$ & $10^{-1}$ & 1 &  52 & 151 \\
$10^{3}$ & $0.1$ & $10^{-2}$ & 1 &  24 & 82 \\
$10^{3}$ & $0.1$ & $10^{-3}$ & 1 &  12 & 41 \\
$10^{3}$ & $0.1$ & $10^{-4}$ & 1 &   8 & 20 \\ \hline
$10^{3}$ & $0.1$ & $10^{-2}$ & 10 &  55 & 149 \\
$10^{3}$ & $0.1$ & $10^{-2}$ & 3 &  36 & 114 \\
$10^{3}$ & $0.1$ & $10^{-2}$ & 1 &  24 & 82 \\
$10^{3}$ & $0.1$ & $10^{-2}$ & 0.3 &  14 & 44 \\ \hline
\end{tabular}
\end{center}
\end{table}

\subsection{Fiducial Disk Model}\label{ssec:fid}
\begin{figure}
\centering
\includegraphics[width = 1.0\hsize,clip]{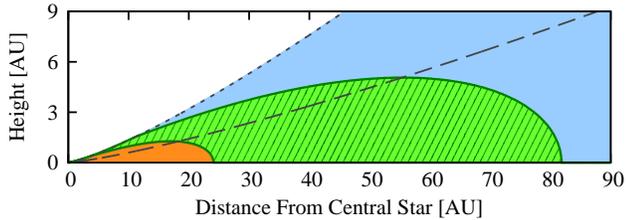}
\caption{Cross-section view of the fiducial protoplanetary disk indicating the 
location of the dead, e-heating and active zones (red, green-shaded, and blue regions, respectively). 
The dashed line shows the gas scale height $H$, while the dotted line shows 
the critical height $H_{\rm ideal}$ below which the MRI criterion in the ideal MHD limit is satisfied
(see Equation \eqref{eq:con-ide-z}).
}
 \label{fig:disk}
\end{figure}

Figure~\ref{fig:disk} shows the two-dimensional (radial and vertical) map
of the dead, active, and e-heating zone in the fiducial disk model.
The MRI criterion in the ideal MHD limit (Equation~\eqref{eq:con-ide}) 
is satisfied at altitudes below $z =  H_{\rm ideal} \approx 2.3H$ (see Equation~\eqref{eq:con-ide-z}).
The region above this height is MRI-stable with the MRI modes suppressed 
by too strong magnetic tension. 
The dead zone is located inside 24 AU from the central star and 
near the midplane where the gas is shielded from ionizing irradiation.
The size of the dead zone for this disk model is consistent with the prediction by \citet{Sano2000Magnetorotation} (see their Figure 7(b)), although their dead zone is slightly
thicker than ours because of the neglect of X-ray ionization.  
We find that the e-heating zone extends from the outer edge of the dead zone 
out to 82 AU from the central star.
This means that MRI turbulence can develop without affected by electron heating 
only in the outermost region of $r \ga 80$ AU.

\begin{figure*}
\epsscale{0.35}
\plotone{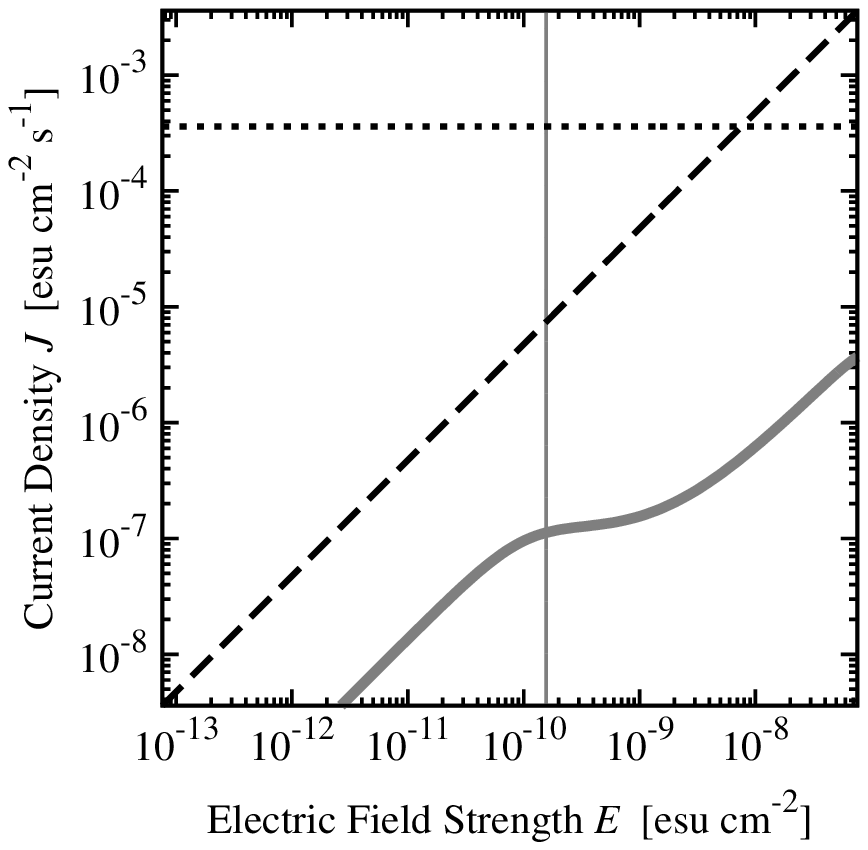}\hspace{3mm}
\plotone{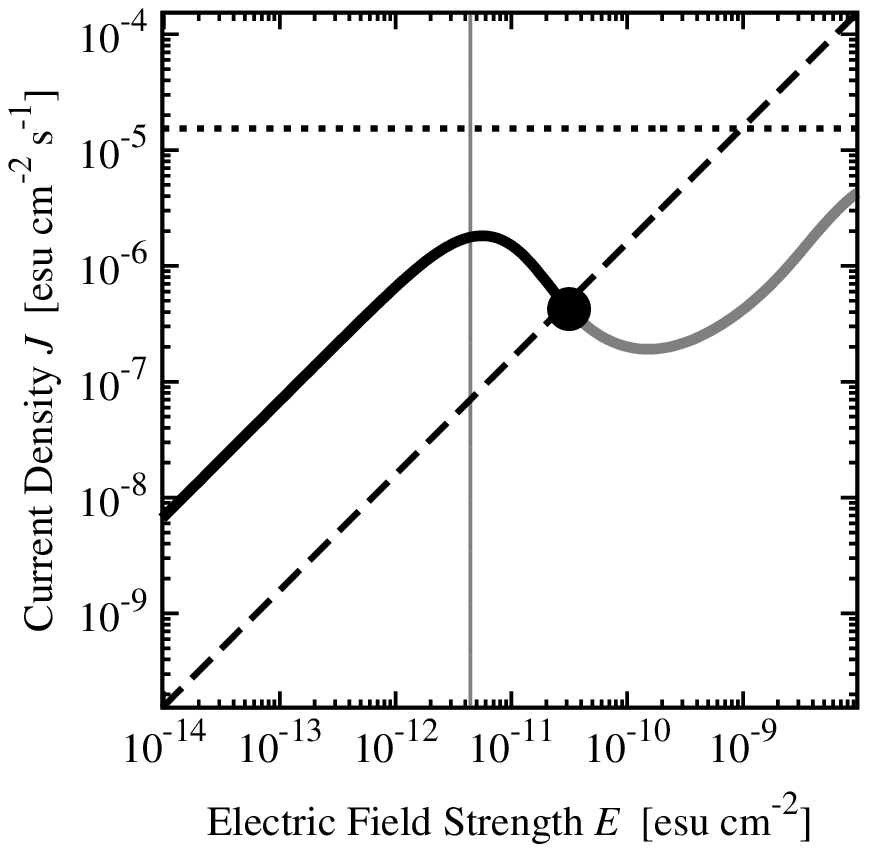}\hspace{3mm}
\plotone{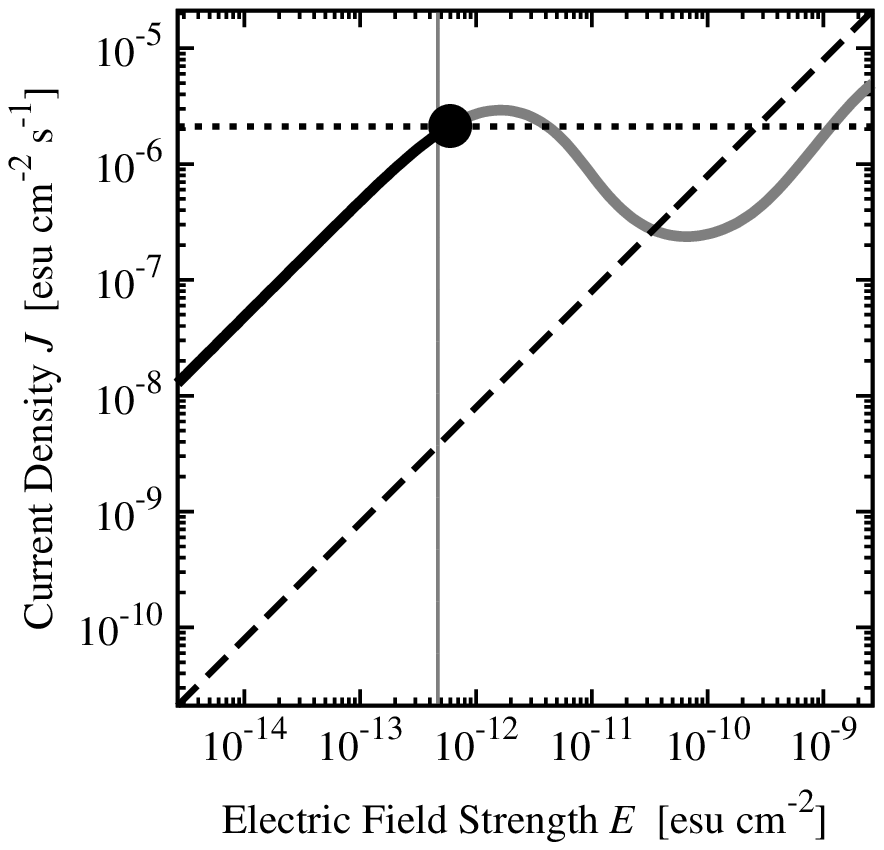}
\caption{
      Relations between the current density $J$ and electric field strength $E$ 
      in the midplane at 15 AU (left panel), 45 AU (middle panel), and 90 AU (right panel), 
      which represent the $J$--$E$ relations in the dead, e-heating, and active zones, respectively. 
      The thick solid line shows the current--field relation $J(E)$, the dotted line
      the maximum current density of MRI, $J_{\rm max}$ (Equation (\ref{eq:Jmax})), 
      the vertical gray solid line the criterion for electron heating, $E_{\rm crit}$ (Equation (\ref{eq:Ecrit})),
      and the dashed line the critical current density $J_{\Lambda = 1}$ below which 
      the MRI decays owing to Ohmic dissipation (Equation \eqref{eq:Jlam}).
      The black dots on the $J$--$E$ relations indicate the saturation points at which either fully developed 
      ($J(E) = J_{\rm max}$) or self-regulated ($J(E) = J_{\rm \Lambda=1}$) MRI turbulence is sustained.
      }
      \label{fig:JEdia}
\end{figure*}
To illustrate how our zoning criteria work in this particular example, 
we plot in Figure~\ref{fig:JEdia} the relation between the current density $J$ and electric field $E$ 
in the midplane at 15 AU, 45 AU and 90 AU,
which represent the dead, e-heating, and active zones, respectively. 
Recall that for fixed $E$, MRI turbulence grows if $J(E) > J_{\Lambda = 1}$ and decays otherwise 
(Equation~\eqref{eq:OC-dis-ex}).
At 15 AU, $J(E)$ falls below $J_{\Lambda = 1}$ for all values of $E$, implying that 
the MRI is unable to grow at this location. 
At 45 AU, the MRI growth condition is satisfied 
during the initial growth stage of $E \ll 10^{-11}~{\rm esu~cm^{-2}}$,
but breaks down {\it before $J$ reaches $J_{\rm max}$} because of the decrease in $J(E)$
due to electron heating.
This implies that MRI turbulence is allowed to grow in the initial stage
but saturates at a level lower than that for fully developed turbulence.   
At 90 AU, $J(E)$ reaches $J_{\rm max}$ before electron heating sets in, 
implying that fully developed MRI turbulence is sustained here. 

In contrast to electron heating, ion heating is found to be negligible at all locations in the fiducial disk model. 
In the e-heating zone, the electric field strength at the saturation point is typically $\la 10^2 E_{\rm crit}$
(see the center and right panels of Figure \ref{fig:JEdia}), 
which is an order of magnitude lower than the field strength required for ion heating, $10^{3} E_{\rm crit}$.

\subsection{Dependence on the Magnetic Field Strength}
\begin{figure*}
\epsscale{1.0}
\plotone{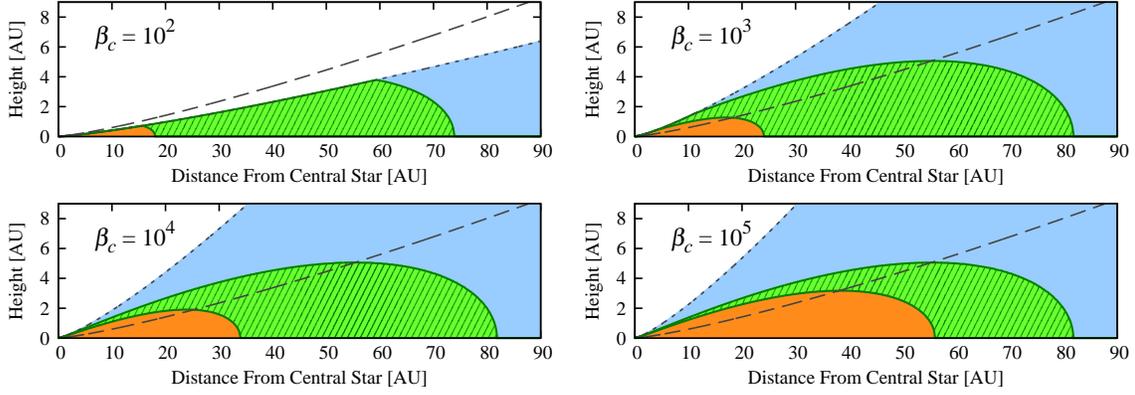}
\caption{
Same as Figure~\ref{fig:disk}, but for different values of the midplane beta $\beta_c$. 
}
\label{fig:disks-beta}
\end{figure*}
Figure \ref{fig:disks-beta} shows how the size of the dead and e-heating zones depend on 
the midplane plasma beta $\beta_c$. 
Recall that a higher $\beta_c$ corresponds to a weaker magnetic field $B$ threading the disk. 
As we increase $\beta_{c}$, the dead zone expands because the Elsasser number 
$\Lambda \propto B^{2}$ decreases. 
On the other hand, we find that the boundary between the e-heating and active zones is less sensitive to 
the choice of $\beta_c$.
As can be inferred from the middle and right panels of Figure~\ref{fig:JEdia}, 
this boundary is approximately determined by the condition that 
the current density $J(E)$ reaches $J_{\rm max}$ at a local maximum lying at $E \approx E_{\rm crit}$.
Since both $E_{\rm crit}$ and $J_{\rm max}$ are independent of $B$ and hence of $\beta_c$,
so is the boundary between the e-heating and active zones.

\subsection{Dependence on the Grain Size and Dust-to-Gas Mass Ratio}\label{ssec:disks-a}
The size and amount of dust grains in disks are important parameters
in the ionization model as they efficiently remove plasma particles from the gas phase.
Obviously, these quantities change as the grains coagulate, settle, or are incorporated by
even larger solid bodies like planetesimals.  
We here explore how the change of these parameters affect 
the size of the dead and e-heating zones.

\begin{figure*}
\epsscale{1.0}
\plotone{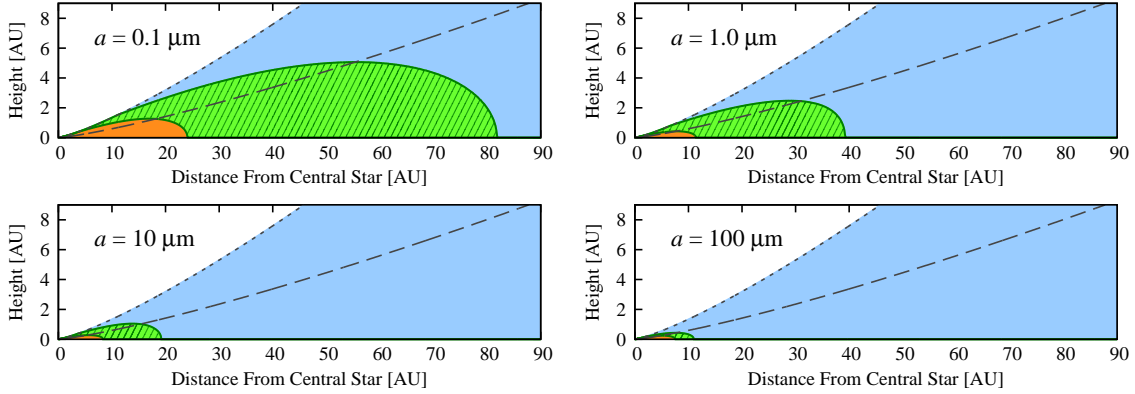}
\caption{
Same as Figure~\ref{fig:disk}, but for different values of the grain size $a$. 
 }
\label{fig:disks-a}
\end{figure*}
To begin with, we show in Figure \ref{fig:disks-a} the location of the dead, active, and e-heating zones
with the dust-to-gas ratio $f_{dg}$ fixed to 0.01 but with the grain size $a$ 
varying between $0.1~{\rm \mu m}$ and $100~{\rm \mu m}$.
We can see that the e-heating zone shrinks with increasing grain size.
On increasing $a$ by a factor of 10, 
the outer radius of the e-heating zone decreases by a factor of $\approx 2$.
Qualitatively, this is simply because the ionization fraction of the gas increases with decreasing 
total surface area of the grains.  
Equation~\eqref{eq:ch-bal-e} shows that the electron abundance $x_e = n_e/n_n$
in equilibrium is inversely proportional to the total surface area of grains per unit volume $4 \pi a^2 n_d$ as long as 
adsorption of plasma particles onto the grains dominate over gas-phase recombination. 
When dust grains aggregate, their total surface area decreases inversely proportional to $a$,
and hence the electron abundance increases linearly with $a$. 
The resulting increase in the electric conductivity causes 
a shift of the $J$--$E$ curve toward higher $J$, enabling the curve to cross the $J= J_{\rm max}$
line at smaller orbital radii.  
We also find that the outer radius of the dead zone decreases at a similar rate to 
that of the e-heating zone when we go from $a = 0.1~{\rm \mu m}$ to $1~{\rm \mu m}$. 
However, the decrease in the dead zone size stops 
beyond this grain size because gas-phase recombination takes over plasma adsorption onto dust grains. 
As a consequence, the e-heating zone becomes narrower and narrower as $a$ increases 
beyond $10~{\rm \mu m}$.

\begin{figure*}
\epsscale{1.0}
\plotone{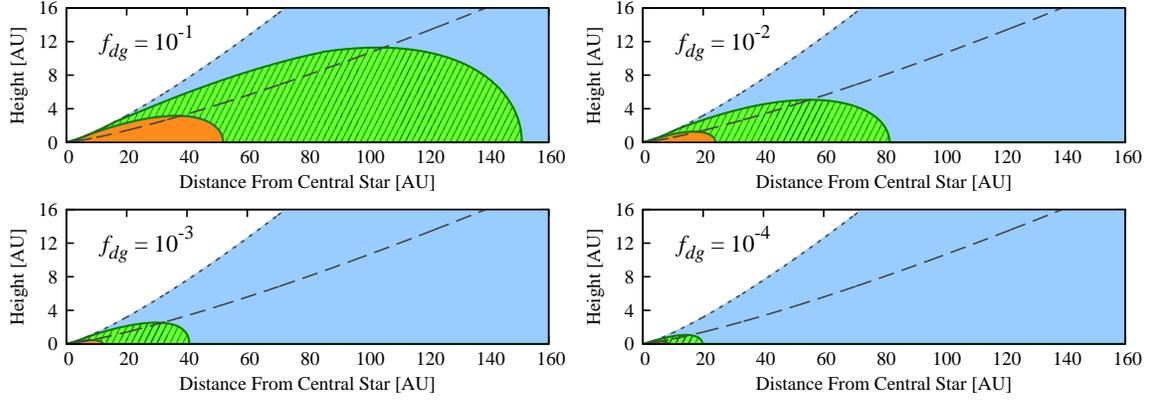}
\caption{
Same as Figure~\ref{fig:disk}, but for different values of the dust-to-gas mass ratio $f_{dg}$. 
}
\label{fig:disks-fdg}
\end{figure*}
\begin{figure*}
\plotone{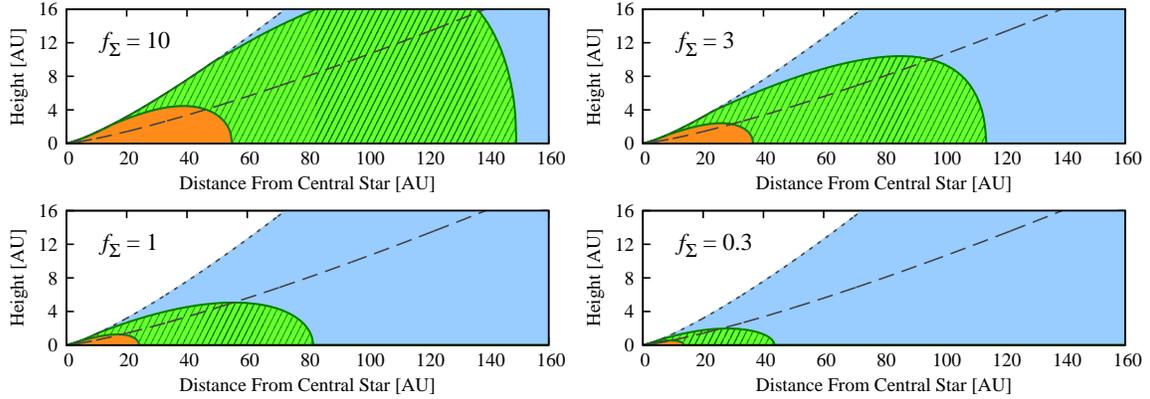}
\caption{
Same as Figure~\ref{fig:disk}, but for different values of the surface density scaling factor $f_\Sigma$. 
}
\label{fig:disks-fS}
\end{figure*}
Decreasing the dust-to-gas mass ratio $f_{dg}$ has a similar effect to increasing the grain radius
because the total surface area of the grains is linearly proportional to $f_{dg}$.
This can be seen in Figure \ref{fig:disks-fdg}, where we show the location 
of the dead and e-heating zones for $a = 0.1~\micron$ with $f_{dg}$ varying 
between $10^{-1}$ and $10^{-4}$. 
We see that the outer radii of the active and e-heating zone decrease by a factor of $\approx 2$ 
when $f_{dg}$ is decreased by a factor of $10$. 
This trend is similar to what we have seen when increasing the grain radius by the same factor.

\subsection{Dependence on the Disk Mass}
Finally, we examine how the size of the e-heating zone depends on the disk mass. 
Figure \ref{fig:disks-fS} shows the location of the e-heating zone for different values of $f_{\Sigma}$.
Here, we fix the dust-to-gas mass ratio $f_\Sigma$ so that 
both the gas and dust densities scale with $f_{\Sigma}$.
We find that the e-heating zone expands toward larger orbital radii and higher altitudes as $f_\Sigma$ increases. 
In the horizontal direction, the expansion is mainly due to the increased amount of dust grains
with increasing $f_\Sigma$. 
As we have explained in \ref{ssec:disks-a}, the ionization fraction of the gas scales inversely 
with $4 \pi a^2 n_d$, and hence with $f_\Sigma$.
Therefore, increasing $f_\Sigma$ by a factor 
has the same effect as increasing $f_{dg}$ by the same factor
as long as the ionization rate $\zeta$ is unchanged 
(which is approximately true at $\sim100~{\rm AU}$ where cosmic rays penetrate down to the midplane). 
This is exactly what we see in Figures~\ref{fig:disks-fdg} and \ref{fig:disks-fS}, 
where the e-heating zone expands to 150 AU when either $f_{dg}$ or $f_\Sigma$
is increased by the factor of 10 from the fiducial value. 
By contrast, the vertical expansion of the e-heating zone is caused by the attenuation of X-rays 
that occurs at higher altitudes with increasing gas column density.

\subsection{Ion heating}\label{ssec:IH}

\begin{figure}
\centering
\includegraphics[width = 0.9\hsize,clip]{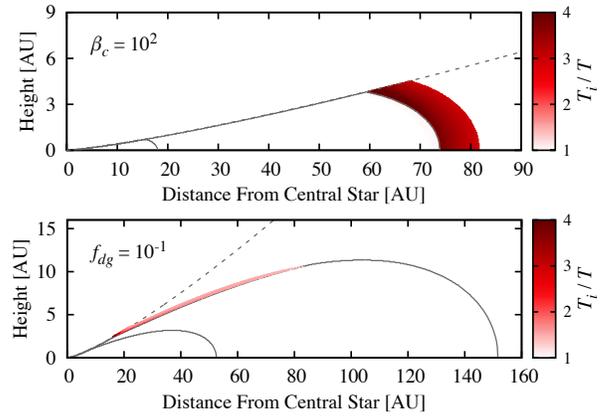}
\caption{
Two-dimensional distribution of the ion temperature $T_i$ for the $\beta_c = 100$ model (upper panel) and $f_{dg} = 0.1$ model (lower panel). The solid lines show the boundary of the e-heating zone, while the dotted lines show $H_{\rm ideal}$.
}
\label{fig:disks-HI}
\end{figure}

We observe ion heating in two cases where $\beta_{c} = 100$ and where $f_{dg} = 0.1$. 
Figure \ref{fig:disks-HI} plots the distribution of the ion temperature $T_i$ in the saturated state for these cases. 
In the case of $\beta_c = 100$ (the upper panel of Figure \ref{fig:disks-HI}), $T_i$ is 3--4 times higher than the temperature in a region slightly outside the e-heating zone.  
In this case, the Elsasser number $\Lambda$ exceeds unity even after electron heating reduces $\Lambda$. 
This allows the electric field strength to reach the critical value for ion heating ($\approx 10^3E_{\rm crit}$) in the vicinity of the e-heating zone. 
In the case of $f_{dg} = 0.1$ (the lower panel of Figure \ref{fig:disks-HI}), ion heating takes place near the upper boundary of the e-heating zone. 
However, the region is very narrow, and the temperature rise is less than $2T$.
Therefore, in this case, ion heating might be practically negligible.

\section{Saturation of Turbulence in the E-heating Zone}\label{sec:Suppress}

\begin{figure*}
\epsscale{0.95}
\plottwo{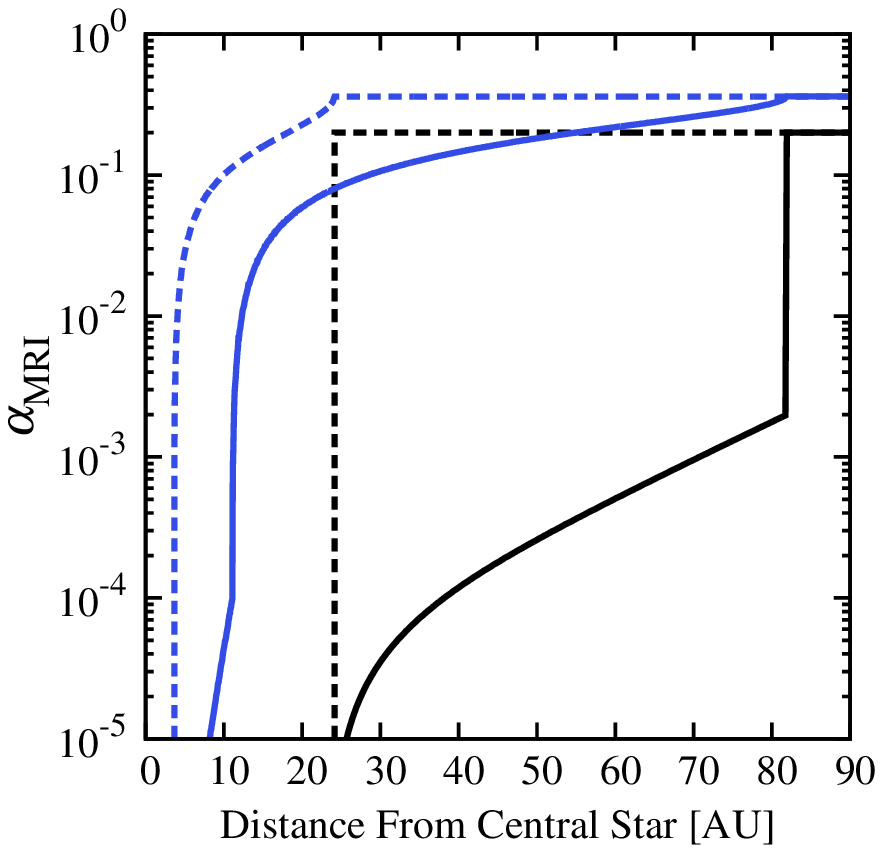}{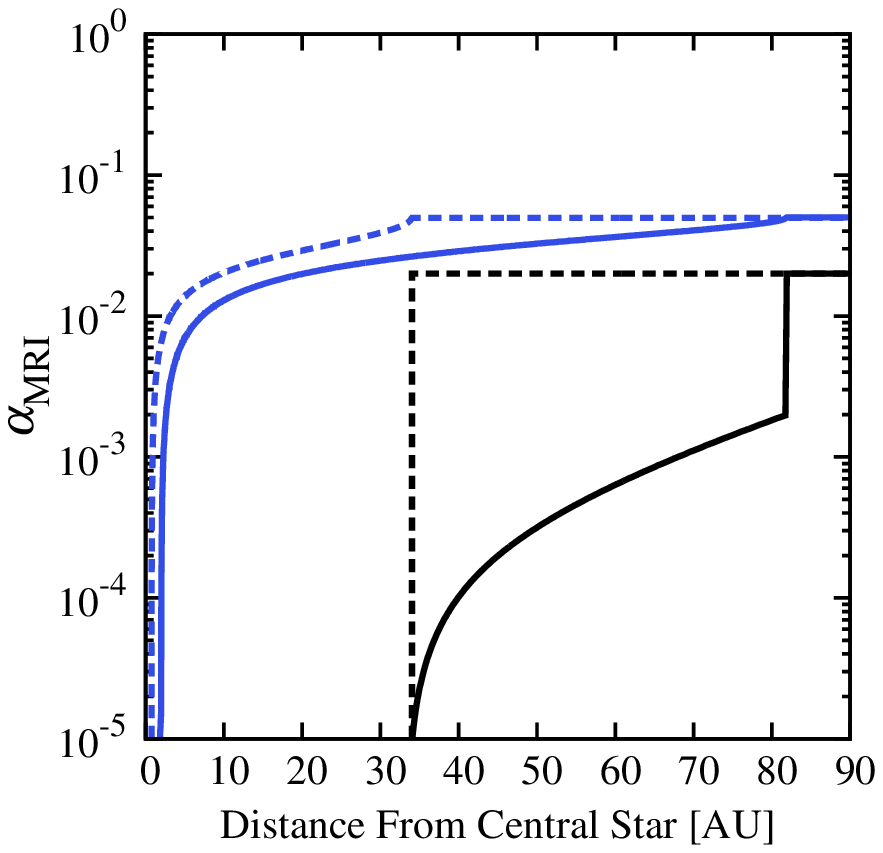}
\caption{
Radial distribution of $\alpha_{\rm MRI}$ (Equation~\eqref{eq:alpha-2}) for the fiducial model (left panel) and $\beta_{c} = 10^{4}$ (right panel).
The solid black line shows $\alpha_{\rm MRI,mid}$ including electron heating on the mid-plane, and the solid blue line shows $\bar{\alpha}_{\rm MRI}$ including electron heating integrated in the z-direction.
The dashed black line shows $\alpha_{\rm MRI,mid}$ without including electron heating on the mid-plane, and the dashed blue line shows $\bar{\alpha}_{\rm MRI}$ without including electron heating integrated in the z-direction.}
\label{fig:alp}
\end{figure*}

We have shown in Section~\ref{sec:Results} that self-regulation of the MRI due to electron heating 
can occur over a large region of protoplanetary disks. 
Then the question arises how strongly the e-heating will suppress the MRI turbulence in the e-heating zones.
This question can only be fully addressed with MHD simulations including magnetic diffusion and electron heating in a self-consistent manner, which is far beyond the scope of this study. 
In this section, we attempt to estimate the saturation level of MRI turbulence from simple scaling arguments.

As usual, we quantify the strength of turbulence with the Shakura--Sunyaev $\alpha$ parameter 
$\alpha = T_{r\phi}/P$, where $P = \rho c_s^2$ is the gas pressure and
$T_{r\phi}$ is the $r\phi$ component of turbulent stress. 
In MRI-driven turbulence, $T_{r\phi}$ is generally dominated by the turbulent Maxwell stress 
$-\delta B_r \delta B_\phi/4\pi$ \citep{Hawley1995Local-Three-dim,Miller2000The-Formation-a},  
where $\delta B_{r}$ and $\delta B_{\phi}$ are the radial and azimuthal components of 
the turbulent (fluctuating) magnetic fields.  
Therefore, we evaluate the $\alpha$ parameter for MRI turbulence as
\begin{equation}\label{eq:alpha}
	 \alpha_{\rm MRI}  \approx  -\frac{\delta B_{r} \delta B_{\phi}}{4 \pi \rho c_s^2}.
\end{equation}
In reality, the Reynolds stress \citep{Fleming2003Local-Magnetohy,Okuzumi2011Modeling-Magnet} 
or the coherent component of the Maxwell stress \citep[e.g.,][]{Turner2008Dead-Zone-Accre,Gressel2011On-the-dynamics} can dominate over the turbulent Maxwell stress at locations where the MRI is significantly suppressed.
However, we do not include these components in our $\alpha_{\rm MRI}$ because they do not reflect the local MRI activity at such locations (see the references above). 

Next we relate the amplitude of turbulent magnetic fields to
the amplitude of the electric current density $J =|{\bm J}|$ using the Ampere's law
${\bm J} = (c/4\pi)\nabla\times \delta{\bm B}$. 
We neglect large-scale, coherent components in ${\bm B}$ since the electric current is inversely 
proportional to the length scale of fields. 
We assume that the magnetic field in MRI-driven turbulence is dominated by 
the azimuthal component $\delta B_\phi$ and varies over a length scale 
$\sim \lambda_{\rm ideal}$, where $\lambda_{\rm ideal}$ is the wavelength of the most unstable MRI modes
already introduced in Equation~\eqref{eq:lambda_ide}.
Then, from the Ampere's law, one can estimate the magnitude of the current density as 
\begin{eqnarray}\label{eq:j-db}
	J &=& \frac{c}{4 \pi} |\nabla \times  \bm{B}| \nonumber \\
      &\approx& \frac{c}{4 \pi} \frac{v_{Az}}{\Omega}\delta B_{\phi} 
      	 = \sqrt{\frac{\rho}{4\pi}}c\Omega \frac{\delta B_{\phi}}{B_{z}},
\end{eqnarray}
where we have replaced the derivative $\nabla$ with wavenumber $2\pi/\lambda_{\rm ideal} = \Omega/v_{Az}$.
If we use the maximum current $J_{\rm max}$ for fully developed MRI turbulence (Equation \eqref{eq:Jmax}), 
Equation~\eqref{eq:j-db} results in a simple scaling relation
\begin{equation}\label{eq:db-rel}
	\frac{\delta B_\phi}{B_z} \approx 10\sqrt{2}\frac{J}{J_{\rm max}}.
\end{equation}
For fully developed MRI turbulence where $J \approx J_{\rm max}$, 
the above equation predicts $\delta B_{\phi}/B_{z} \sim 10$, 
in agreement with the results of MHD simulations \citep[e.g.,][]{Hawley1995Local-Three-dim,Sano2004Angular-Momentu}.

Now let us consider situations where e-heating is so effective that the growth of the MRI 
is saturated at $J \approx J_{\rm \Lambda=1} \ll J_{\rm max}$. 
Assuming $\delta B_{z} \la B_{z0}$ for this case, we have
\begin{equation}\label{eq:By-Delta}
	\delta B_\phi \approx 10\sqrt{2}B_{z0} \frac{J}{J_{\rm max}}.
\end{equation}
This equation predicts the amplitude of $\delta B_\phi$ as a function of $B_{z0}$ and 
$J/J_{\rm max}$. 
MHD simulations show that $\delta B_r \approx -(0.4\dots0.6)\delta B_{\phi}$ in MRI turbulence 
\citep{Hawley1995Local-Three-dim,Sano2004Angular-Momentu}. 
Assuming that this scaling also holds in our case, 
we have $\delta B_{r} \delta B_{\phi} \approx -100B_{z0}^2({J}/{J_{\rm max}})^2$.
Finally, substituting this into Equation~\eqref{eq:alpha}, we obtain the scaling relation between 
$\alpha_{\rm MRI}$ and $J/J_{\rm max}$, 
\begin{eqnarray}\label{eq:alpha-2}
	\alpha_{\rm MRI} &\approx& 
	\frac{  100 B_{0}^{2} }{4 \pi \rho c_s^2} \left(\frac{J}{J_{\rm max}}\right)^2 \nonumber\\
	&\approx& 0.2\Pf{\beta_{0}}{1000}^{-1}\left(\frac{J}{J_{\rm max}}\right)^2,
\end{eqnarray}
where $\beta_{0} \equiv 8\pi \rho c_s^2/B_{z0}^2 = \beta_c \exp{(-z^{2}/2H^{2})}$
is the plasma beta (not necessarily at the midplane) associated with the net vertical field $B_{z0}$.
Formally, the derivation leading to Equation~\eqref{eq:alpha-2} breaks down when 
MRI is so active that $\delta B_z \gg B_{z0}$ and $J \approx J_{\rm max}$. 
Nevertheless, we find that Equation~\eqref{eq:alpha-2} reproduces the results of ideal MHD simulations 
with a reasonably good accuracy. 
Equation~\eqref{eq:alpha-2} predicts that $\alpha_{\rm MRI} \approx 2$
for $\beta_0 = 10^2$ and $\alpha_{\rm MRI} \approx 0.02$ for $\beta_0 = 10^2$ 
when $J = J_{\rm max}$. 
These are consistent with the results of isothermal simulations by \citet{Sano2004Angular-Momentu} 
showing that the Maxwell component of $\alpha$ is $\sim 1$ for $\beta_0 = 10^2$ and $\sim 0.01$ 
for $\beta_0 = 10^4$ (see their Table 2, column (10)). 
Therefore, we will apply Equation~\eqref{eq:alpha-2} to both the e-heating zone and active zone.

The left panel of Figure~\ref{fig:alp} show the radial distribution of $\alpha_{\rm MRI}$ for the fiducial disk model 
predicted from Equation~\eqref{eq:alpha-2}. 
Here we plot the midplane value $\alpha_{\rm MRI,mid} \equiv \alpha_{\rm MRI}(z=0)$ 
and the density-weighted average in the vertical direction, 
\begin{equation}\label{eq:alpha_bar}
	\bar{\alpha}_{\rm MRI} \equiv \frac{\int^{H_{\rm ideal}}_{-H_{\rm ideal}} \alpha_{\rm MRI}(z') \rho(z') dz' }{\Sigma},
\end{equation}
where we have assumed $\alpha_{\rm MRI} = 0$ in the magnetically dominated atmosphere
at $|z| > H_{\rm ideal}$. 
The former quantity measures the MRI activity at the disk midplane, while 
the latter quantity is more closely related to the vertically integrated mass accretion rate \citep{Suzuki2010Protoplanetary-}.
For the fiducial disk model, we find that $\alpha_{\rm MRI, mid} \sim 10^{-5}$ and $10^{-3}$ 
at the inner and outer edge of the e-heating zone (20 AU and 80 AU), respectively. 
These values are more than two orders of magnitude lower than the value 
$\alpha_{\rm MRI,mid} = 0.2$ in the active zone ($r \ga 80~{\rm AU}$).
This implies that the MRI is ``virtually dead'' deep inside the e-heating zone. 
We also find that $\alpha_{\rm MRI,mid}$ changes discontinuously at the boundary between the e-heating and active zones.
The reason is that when the saturated state changes at the point, $J/J_{\rm max}$ also changes from unity to one order of magnitude because of the N-shaped current--field relation (see middle and right panels of Figure \ref{fig:JEdia}). 
The vertical average $\bar{\alpha}_{\rm MRI}$ decreases more slowly with decreasing $r$, 
because the upper layer of the disk remains MRI-active (see Figure \ref{fig:disk}).
This picture is qualitatively similar to the classical layered accretion model of \citet{Gammie1996Layered-Accreti}.
In right panel of Figure~\ref{fig:alp}, we also plot the radial distribution of $\alpha_{\rm MRI,mid}$ and $\bar{\alpha}_{\rm MRI}$ for a disk with $\beta_{c} = 10^{4}$.
We find that $\alpha_{\rm MRI, mid}$ in e-heating zone is almost unchanged from the fiducial disk.
The reason is that increase of $(J/J_{\rm max})^{2} \approx 10$ cancels  out the depletion of $\beta_{c}^{-1}\approx 10^{-1}$ in Equation \eqref{eq:alpha-2}.
Therefore, $\alpha_{\rm MRI, mid}$ remains low saturation level.

In summary, our simple estimate predicts that MRI turbulence can be significantly suppressed in the e-heating zone.
In this sense, the e-heating zone acts as a extended dead zone.
However, our estimate relies on the hypothetical scaling between the and turbulent Maxwell stress
and $J/J_{\rm max}$, which is as yet justified by MHD simulations.\footnote{
However, there are some support for Equation~\eqref{eq:alpha-2} from MHD simulations including ambipolar diffusion, 
not Ohmic dissipation.  
\citet{Bai2011Effect-of-Ambip} reported the Maxwell component of $\alpha$ (their Table 2)
and the cumulative probability distribution of $J$ (Figure 6) for three simulation runs
with $\beta_0 = 400$ and with different values of ambipolar diffusivity. 
Their results show that $\alpha_{\rm Maxwell} \approx 0.17$, 0.029, and $0.0041$
for models with $J/J_{\rm max} \approx 1, 0.3$, and 0.1 (median values), respectively.
These are consistent with Equation~\eqref{eq:alpha-2} predicting 
that $\alpha_{\rm MRI} \approx 0.5$, $0.045$, and $0.005$ for these values of $J/J_{\rm max}$.
} 
In order to test our prediction, we will perform resistive MHD simulations including electron heating
in future work.

\section{Charge Barrier against Dust Growth in the E-heating Zone}\label{sec:CB}
So far we have focused on the role of electron heating on the saturation of MRI turbulence. 
As pointed out by \citetalias{Okuzumi2015The-Nonlinear-O}, electron heating also has
an important effect on the growth of dust grains.
In an ionized gas, dust grains tend to be negatively charged because 
electrons collide and stick to dust grains more frequently than ions. 
The resulting Coulomb repulsion slows down the coagulation of the grains through Brownian (thermal) motion.  
This ``charge barrier'' is also present in weakly ionized protoplanetary disks, 
in which dust grains tend to be charged as in a fully ionized gas
when their size is larger than 1 $\micron$ \citep{Okuzumi2009Electric-Chargi,Matthews2012Charging-and-Co}. 
The important role of electron heating in this context is that 
heating electrons further promote the negative charging of the grains,
because the grain charge in a plasma is linearly proportional to the electron temperature \citep[e.g.,][]{Shukla2002Introduction-to}.
In this section, we explore how this affects dust coagulation in the e-heating zone.  

For simplicity, let us assume that dust grains have the single radius $a$ and charge $Z$.
The grains can collide with each other if the condition
\begin{equation}
	{\cal E}_{\rm col} > {\cal E}_{\rm elc}
\end{equation}
is satisfied \citep{Okuzumi2009Electric-Chargi}. 
Here, ${\cal E}_{\rm col}$ is the kinetic energy of the relative motion of two colliding grains, and 
\begin{equation}
	{\cal E}_{\rm elc} \approx \frac{(eZ)^{2}}{2a}
\end{equation}
is the Coulomb repulsion energy of the grains just before contact. 
We focus on small dust grains near the midplane and assume that 
the relative motion is dominated by Brownian motion and turbulence-induced motion. 
Then, the kinetic energy of relative motion can be expressed as 
\begin{equation}\label{eq:}
	{\cal E}_{\rm col} = {\cal E}_{\rm Brown} + {\cal E}_{\rm turb},
\end{equation}
where ${\cal E}_{\rm Brown}$ and ${\cal E}_{\rm turb}$ are the kinetic energy of
Brownian motion and turbulence-induced motion, respectively. 
Brownian motion is the thermal motion of grains, 
and ${\cal E}_{\rm Brown}$ is approximately expressed as
\begin{equation}\label{eq:}
	{\cal E}_{\rm Brown} \approx \frac{1}{2}\mu u_{\rm th}^{2} ,
\end{equation}
where the thermal velocity of grains $u_{\rm th}$ is expressed as $u_{\rm th} = \sqrt{8kT/\pi m}$ and the reduced mass of grains $\mu$ is expressed as $\mu = m^{2}/(m+m)=m/2$.
The relative energy of turbulence-induced motion is expressed as 
\begin{equation}\label{eq:}
	{\cal E}_{\rm turb} \approx \frac{1}{2}\mu (\Delta u_{\rm turb})^{2},
\end{equation}
where $\Delta u_{\rm turb}$ is the relative velocity of the grains excited by turbulence.
For small grains, $\Delta u_{\rm turb}$ is approximately given by \citep{Weidenschilling1984Evolution-of-gr,Ormel2007Closed-form-exp}
\begin{equation}\label{eq:uturb}
	\Delta u_{\rm turb} \approx \sqrt{\alpha_{\rm disp}}{\rm Re}^{1/4} c_{s} \Omega \tau_s,
\end{equation}
where $\alpha_{\rm disp} \equiv \avg{\delta v^{2}}/c_s^2$ is the velocity dispersion of the gas $\avg{\delta v^{2}}$
normalized by $c_s^2$,  ${\rm Re}$ is the Reynolds number of turbulence, and 
\begin{equation}\label{eq:}
	\tau_s = \rho_\bullet a/(\sqrt{8/\pi} c_s\rho)	
\end{equation}
is the stopping time of the grains
(we have adopted Epstein's drag law for $\tau_s$). 
The Reynolds number is expressed as ${\rm Re} = \alpha_{\rm disp}  c_{s}^{2}\Omega^{-1} / \nu_{\rm mol}$, 
where $\nu_{\rm mol}$ is the molecular viscosity.
We estimate $\alpha$ with and without electron heating, 
using Equation~\eqref{eq:alpha-2} presented in Section \ref{sec:Suppress}.
Turbulence dominates the collisional energy when $\alpha_{\rm disp}$ is high and/or $a$ is large. 
For the moment, we simply assume $\alpha_{\rm disp} = \alpha_{\rm MRI}$, 
where $\alpha_{\rm MRI}$ is the normalized local Maxwell stress introduced 
in Equation~\eqref{eq:alpha}. This assumption holds when the Reynolds stress 
in the e-heating zone is comparable to the Maxwell stress.
In reality, the Reynolds stress in the e-heating zone might be higher 
than the Maxwell stress for a reason discussed later. 
Therefore, the estimate of ${\cal E}_{\rm turb}$ presented here should be taken as a lower limit.

To obtain $Z$ and $\alpha_{\rm MRI}$, we calculate the ionization fraction (Section \ref{ssec:ionz-bal}), determine the turbulent state (Section \ref{ssec:DetTS}), and estimate the MRI-turbulent viscosity (Section \ref{sec:Suppress}) with changing grain radius $a$ at a location.
We then obtain ${\cal E}_{\rm col}$ and ${\cal E}_{\rm elc}$ by above-mentioned method.
It should be noted that grains have single size and changing grain radius means changing the size of all grains at the location.
Thus, the turbulent state at the location also depends on $a$.

\begin{figure}
\includegraphics[width = 0.9\hsize,clip]{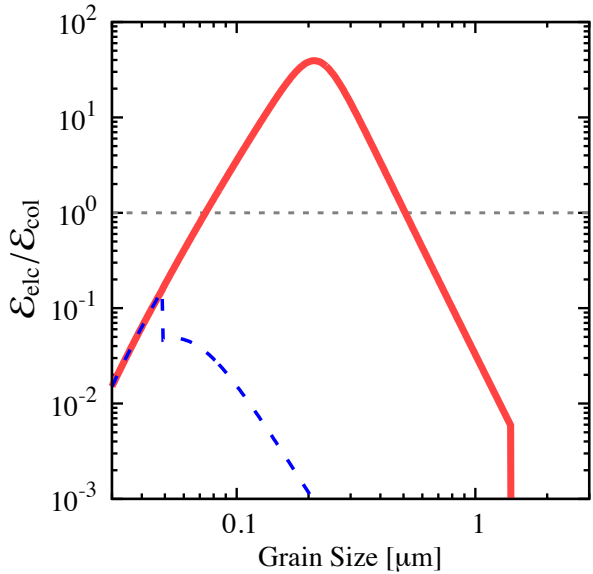}
\caption{
Effectiveness of the charge barrier against grain growth as a function of the grain size at the midplane 35 AU in the fiducial model.
The solid line (red) shows ${\cal E}_{\rm elc}/{\cal E}_{\rm col}$ including electron heating, and the dashed line (blue) shows  ${\cal E}_{\rm elc}/{\cal E}_{\rm col}$ without including electron heating. 
The horizontal dotted line shows ${\cal E}_{\rm elc}/{\cal E}_{\rm coll} = 1$, above which a strong Coulomb repulsion between the grains suppresses their mutual collision cross section.
Here it is assumed that $\alpha_{\rm disp} (= \avg{\delta v^{2}}/c_s^2)$ is equal to $\alpha_{\rm MRI}$,
the normalized local Maxwell stress given by Equation~\eqref{eq:alpha} (but see also Figure~\ref{fig:eel-2}). 
}
\label{fig:eel}
\end{figure}

In Figure \ref{fig:eel}, we plot the ratio ${\cal E}_{\rm elc}/{\cal E}_{\rm col}$ as a function of $a$ at 35 AU in the midplane for the fiducial disk model.
The ratio quantifies the effectiveness of the charge barrier: the collisional cross section of two equally charged grains is
significantly suppressed when ${\cal E}_{\rm elc}/{\cal E}_{\rm col} \gg 1$. 
We find that electron heating significantly enhances the charge barrier for submicron-sized grains.
If electron heating is not included, this location belong to the dead zone and the active zone with grain size being $\la 0.05~{\rm \mu m}$ and $ \ga 0.05~{\rm \mu m}$, respectively.
In this case, ${\cal E}_{\rm elc} / {\cal E}_{\rm col}$ is much lower than unity in all $a$. 
Thus we can conclude that dust grains at this location can grow without the charge barrier.
On the other hand, if electron heating is included, this location belongs to the e-heating zone when $0.05~{\rm \mu m} \la a \la 1.4~{\rm \mu m}$ (see also Figure~\ref{fig:disks-a}).
In the e-heating zone, grains are charged by heated electrons, leading to increase of ${\cal E}_{\rm elc}$, and MRI turbulence as collisional source is well suppressed, leading to decrease of ${\cal E}_{\rm turb}$.
Consequently, ${\cal E}_{\rm elc}/{\cal E}_{\rm col}$ is larger than unity when $0.08~{\rm \mu m} \la a \la 0.5~{\rm \mu m}$. 
In particular, ${\cal E}_{\rm elc}/{\cal E}_{\rm col}$ takes its maximum value of 40 at $a = 0.2~ \mathrm{\mu m} $ corresponding to ${\cal E}_{\rm Brown} = {\cal E}_{\rm turb}$.
Both the suppression of turbulence and grain charge would enhance the charge barrier.

There are at least two mechanisms that could drive further growth of dust in the e-heating zone.
One is vertical turbulent mixing of dust particles as already 
pointed out by \citet{Okuzumi2011bElectrostatic-B}.
In general, the charge barrier is less significant at higher altitudes where 
dust particles have a higher collision energy 
due to vertical settling (and due to if MRI is active there).
Electron heating, which was not considered by \citet{Okuzumi2011bElectrostatic-B}, 
does not change this picture because it is also ineffective at high altitudes. 
Micron-sized grains in the e-heating zone can easily be lifted up to such high altitudes 
if only weak turbulence is present there
\citep[][see also dust scale height $H_{d}$ in Section \ref{ssec:dussed}]{Turner2010Dust-Transport-}. 
The lifted grains are allowed to collide and grow there until they fall back to the e-heating zone.
In this way, small grains in the e-heating zone are able to continue growing
on a timescale much longer than vertical diffusion timescale.  
\citet{Okuzumi2011bElectrostatic-B} showed that the charge barrier is overcome on a timescale of 
$10^5$--$10^6$ yr, but they did not consider the amplification of grain charging due to electron heating.
How much the growth is delayed in the presence of electron heating should be studied in future work.

\begin{figure}
\includegraphics[width = 0.9\hsize,clip]{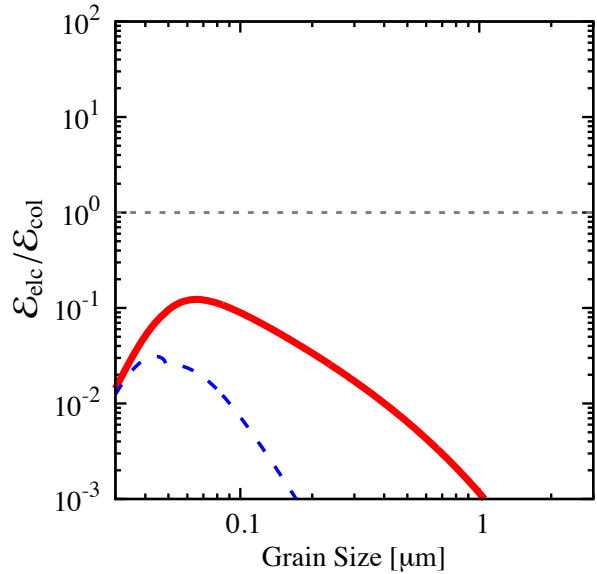}
\caption{
Same as Figure~\ref{fig:eel}, but we here evaluate $\alpha_{\rm disp} = \avg{\delta v^{2}}/c_s^2$ 
using Equation~\eqref{eq:dv2} assuming that sound waves propagate from upper MRI active layers to the midplane as it is observed for the conventional dead zone. 
}
\label{fig:eel-2}
\end{figure}

Another potentially important mechanism is dust stirring by random sound waves.
It is known that the Reynolds stress in a dead zone exceeds 
the Maxwell stress because of sound waves propagating from upper MRI-active layers
\citep[e.g.,][]{Fromang2006Dust-settling-i,Turner2010Dust-Transport-,Okuzumi2011Modeling-Magnet}.
If this is also the case for our e-heating zone, the assumption $\alpha_{\rm disp} = \alpha_{\rm MRI}$
would significantly underestimate the particle collision energy in the e-heating zone.
To estimate this effect, we now calculate $\alpha_{\rm disp}$ using an empirical formula 
for the gas velocity dispersion in the dead zone \citep{Okuzumi2011Modeling-Magnet}, 
\begin{equation}\label{eq:dv2}
	\avg{\delta v^{2}} \approx 0.78 \bar{\alpha}_{\rm MRI} c_{s}^{2} \exp{\Pf{z^{2}}{2H^{2}}},
\end{equation}
where $\bar{\alpha}_{\rm MRI}$ is the density-weighted vertical average of $\alpha_{\rm MRI}$ 
defined by Equation~\eqref{eq:alpha_bar}.
Equation~\eqref{eq:dv2} expresses the amplitude of random sound waves inside a dead zone.
Figure \ref{fig:eel-2} shows ${\cal E}_{\rm elc} / {\cal E}_{\rm col}$ in this case and is obtained in the same way as in Figure \ref{fig:eel} 
but we here use Equation~\eqref{eq:dv2} for $\alpha_{\rm disp}$ in Equation~\eqref{eq:uturb}. 
The use of Equation~\eqref{eq:uturb} for the sound wave-driven collision velocity assumes
that the time correlation of the waves' velocity fluctuations exponentially decays on the timescale of $\Omega^{-1}$ as in the Kolmogorov turbulence.
We find that ${\cal E}_{\rm elc} / {\cal E}_{\rm col}$ now falls below unity at all grain sizes. 
Thus, sound waves traveling from MRI-active layers, if they exist, could help dust overcome the charge barrier in the e-heating zone.
However, the argument made here is not conclusive because the induced
collision velocity depends on the assumed time correlation function,
or equivalently power spectrum, of the random sound waves. If the
power spectrum of the waves has only a small amplitude at high
frequencies (to which small dust particles are sensitive) compared to
the turbulent spectrum, the wave-induced collision velocity would be
lower than given by Equation~\eqref{eq:uturb}. The spectrum of velocity
fluctuations in the e-heating zone should be studied in future MHD simulations.

\section{Discussion}\label{sec:Discussion}

\subsection{Dust Diffusion }\label{ssec:dussed}

We have assumed so far that the dust-to-gas mass ratio is vertically constant. 
This assumption breaks down when dust particles settle toward the midplane. 
If this is the case, the dust-to-gas ratio would decrease at high altitudes,
and consequently the e-heating zone would shrink in the vertical direction 
as expected from Figure~\ref{fig:disks-fdg}.

However, as we will show below, dust settling is negligible even in the e-heating zone  
because even weak turbulence is able to diffuse small grains to high altitudes. 
\citet{Youdin2007Particle-stirri} analytically derived dust scale height $H_{d}$ in the sedimentation-diffusion equilibrium.
If the particle stopping time $ \tau_s$ is much smaller than the Keplerian timescale $\Omega^{-1}$, which is true for small particles, the dust scale height can be approximately written as 
\begin{equation}\label{eq:Hd}
H_{d} \approx H \pr{ 1 + \frac{\mathrm{St}}{ \alpha_{{\rm disp, } z} }  }^{-1/2}	,
\end{equation}
where $\mathrm{St} = \tau_{s} \Omega$ is the so-called Stokes number
and $\alpha_{{\rm disp, } z} = \avg{ \delta v_{z}^{2}  } / c_{s } ^{2}$
is the vertical component of the velocity dispersion normalized by $c_{s}^{2}$.
Equation~\eqref{eq:Hd} implies that dust settling takes place ($H_{d} < H$)
when ${\rm St} > \alpha_{{\rm disp, } z}$.  
Under the disk model employed in this study, $ \mathrm{St} $ can be expressed as
\begin{equation}\label{eq:St}
 \mathrm{St} =  3 \times 10^{-8} \Pf{a}{0.1 ~\mathrm{\mu m} } f_{\Sigma}^{-1} \Pf{r}{1 ~ \mathrm{AU} }^{3/2} \exp{\Pf{z^2}{2H^2}}.
\end{equation}
Therefore, for $a = 0.1~\micron$, dust settling in the e-heating zone ($r \sim 10$--$100 \,\mathrm{AU} $) 
occurs only if  $\alpha_{{\rm disp, } z} \la 10^{-5}$--$10^{-6}$. 
In the e-heating zone, $\alpha_{\rm MRI} \sim 10^{-5}$--$10^{-3}$ at the midplane  (see Figure~\ref{fig:alp}), 
and therefore we may safely neglect dust settling 
even if the Reynolds stress is as small as the Maxwell stress ($\alpha_{{\rm disp, } z} \sim \alpha_{\rm MRI}$).
A larger $a$ does not change this conclusion, 
because we then would have a higher $\alpha_{\rm MRI}$ or the e-heating zone would vanish.

\subsection{Effects of Grain Size Distribution and Porosity}
We have characterized dust grains with a single particle size $a$ assuming that the size distribution of dust grains is narrow.
Under this assumption, the e-heating zone covers only a small part of protoplanetary disks when the particles grow to millimeter sizes (see Figure \ref{fig:disks-a}).
However, caution is required in applying our results to more general cases where particles have a size distribution.
In such cases, the smallest grains tend to dominate the total surface area of dust 
(which controls the ionization balance), whereas the largest grains tend to dominate 
the total mass of dust, simply because smaller grains have a larger area-to-mass ratio. 
Therefore, it is not obvious what the typical particle size is in these cases. 

Here we discuss more quantitatively how we can apply the results of single-size calculations to
cases with a size distribution. 
Let us assume that the particle size distribution is given by the power-law form 
\begin{equation}\label{eq:distr}
\frac{d n_{d}}{d a} = \frac{3 \rho f_{dg}}{8 \pi \rho_{\bullet} \sqrt{a_{\rm max}}} a^{-3.5}
\end{equation}
with $a_{\rm min} < a < a_{\rm max}$ ($a_{\rm min} \ll a_{\rm max}$), 
where $d n_{d} /d a$ is the number density of dust particles per unit particle radius,
and $a_{\rm min}$ and $a_{\rm max}$ are the minimum and maximum particle sizes, respectively.  
The distribution is normalized so that that the total particle mass density 
$\int m_{d} ({d n_{d} }/{da})da$ becomes equal to $\rho f_{dg}$.
Equation~\eqref{eq:distr} applies when the particle size distribution is determined by fragmentation cascade \citep{Dohnanyi1969Collisional-Mod} and is also known to reproduce the size distribution of interstellar dust grains \citep{Mathis1977The-size-distri}.
The quantity we are interested in is the total surface area of the particles 
as it mainly determines the ionization balance in a gas--dust mixture \citep[e.g.,][]{Sano2000Magnetorotation}.  
This can be calculated as
\begin{equation}\label{eq:surface}
	\int_{a_{\rm min}}^{a_{\rm max}} 4\pi a^{2} \frac{d n_{d} }{da}da \approx \frac{3\rho f_{dg}}{\rho_{\bullet}} \frac{1}{\sqrt{a_{\rm min} a_{\rm max}}}.
\end{equation}
Note that the factor $1/\sqrt{a_{\rm min}}$ comes from the fact that the integration in Equation~\eqref{eq:surface} is dominated by the smallest particles 
(because $a^2 ({d n_{d} }/{da})da \propto d(a^{-0.5})$), 
whereas the factor $1/\sqrt{a_{\rm max}}$ 
from the fact that the total mass is dominated by the largest particles. 
By contrast, if all dust particles have a single size $a_{\rm single}$,  
their total surface area is 
$4 \pi a_{\rm single}^{2} n_{d{\rm ,single}} =  3\rho f_{dg} / (\rho_{\bullet} a_{\rm single})$.
Comparing this with Equation~\eqref{eq:surface}, 
we find that the total surface area of particles whose size distribution is given by Equation~\eqref{eq:distr}
is equal to that of single-size particles if
\begin{equation}\label{eq:a_single}
	a_{\rm single} = \sqrt{a_{\rm min} a_{\rm max}}.
\end{equation}  
Since the total surface area approximately determines the ionization state, 
Equation~\eqref{eq:a_single} may be used to generalize the results presented in this study to 
the cases where the particle size distribution obeys Equation~\eqref{eq:distr}.

Observations of millimeter dust emission from protoplanetary disks suggest 
that the largest dust particles in the disk have a size of centimeters \citep[e.g.,][]{Testi2003Large-grains-in,Natta2004A-search-for-ev,Rodmann2006Large-dust-part,Ricci2010Dust-properties}.
Assuming $a_{\rm max} = 1  ~\mathrm{cm} $ and $a_{\rm min}= 0.1  ~\mathrm{\mu m} $,
we obtain $a_{\rm single} = 30  ~\mathrm{\mu m} $.  
In this case, we expect from Table \ref{tbl:beta} that the e-heating zone extends to $\sim 15~ \mathrm{AU}$.
Thus, even if cm-sized grains exist in protoplanetary disks and the total mass of grains is dominated by such large grains, the e-heating zone can be present in the disks. 

For the same reason, large dust particles can alone provide a large
e-heating zone
if the dust particles are highly fluffy aggregates of tiny grains.
\citet{Okuzumi2009Electric-Chargi} showed that the ionization balance
is insensitive to the particle radius
when the fractal dimension is $\approx 2$, for which the total surface
area of the aggregates is approximately
conserved during the aggregation process.

\subsection{Hall Effect and Ambipolar Diffusion}\label{ssec:Ha+Am}
The plasma heating model employed in this study neglects the effects of magnetic 
fields on the motion of plasma particles. 
In terms of non-ideal magnetohydrodynamics, 
this is equivalent to neglecting ambipolar diffusion and Hall effect \citep[see, e.g.,][]{Wardle1999The-Balbus-Hawl}. A full treatment of these non-Ohmic effects introduces to the model additional complexities arising from the relative angle between the magnetic and electric fields (Okuzumi, Mori, \& Inutuska, in prep.), which is beyond the scope of this paper.  In this subsection, we only briefly discuss how plasma heating and these non-ideal MHD effects could affect each other.

\begin{table*}
\begin{center}
\caption{$\mathrm{Am}$ and Ion Abundance $x_{i}$ in E-heating Zone for Various Parameter Sets}
\label{tbl:Am}\smallskip
\begin{tabular}{cccccccc}
\hline \hline
$\beta_{c}$ & $a$ ($\rm \mu m$)& $f_{dg}$ & $f_{\Sigma}$  & \multicolumn{2}{c}{$\mathrm{Am}$ in e-heating zone }& \multicolumn{2}{c}{$x_{i}$ in e-heating zone }\\
                     &                                   &                 &                           &    Inner edge &  Outer edge                                                 &  Inner edge &  Outer edge                                 \\ \hline
$10^{2}$ & $0.1$ & $10^{-2}$ & 1 &  0.14 & 0.56  & 1.9$\times 10^{-12}$ & 4.7$\times 10^{-11}$\\
$10^{3}$ & $0.1$ & $10^{-2}$ & 1 &   0.17 & 0.62 & 3.4$\times 10^{-12}$ & 5.9$\times 10^{-11}$\\
$10^{4}$ & $0.1$ & $10^{-2}$ & 1 &   0.23 & 0.62 & 7.4$\times 10^{-12}$ & 5.9$\times 10^{-11}$  \\
$10^{5}$ & $0.1$ & $10^{-2}$ & 1 &   0.41 & 0.62 & 2.5$\times 10^{-11}$ & 5.9$\times 10^{-11}$ \\ \hline
$10^{3}$ & $0.1$ & $10^{-2}$ & 1 &   0.17 & 0.62 & 3.4$\times 10^{-12}$ & 5.9$\times 10^{-11}$  \\
$10^{3}$ & $1$   & $10^{-2}$ & 1 &   0.21 & 0.72 & 1.7$\times 10^{-12}$ & 2.8$\times 10^{-11}$\\
$10^{3}$ & $10$  & $10^{-2}$ & 1 &    0.43 & 0.82 & 2.3$\times 10^{-12}$ & 1.3$\times 10^{-11}$\\
$10^{3}$ & $100$ & $10^{-2}$ & 1 &   0.54 & 0.72 & 2.6$\times 10^{-12}$ & 5.6$\times 10^{-12}$\\ \hline
$10^{3}$ & $0.1$ & $10^{-1}$ & 1 &   0.16 & 0.57  & 8.5$\times 10^{-12}$ & 1.2$\times 10^{-10}$\\
$10^{3}$ & $0.1$ & $10^{-2}$ & 1 &   0.17 & 0.62  & 3.4$\times 10^{-12}$ & 5.9$\times 10^{-11}$ \\
$10^{3}$ & $0.1$ & $10^{-3}$ & 1 &  0.24 & 0.71  & 2.1$\times 10^{-12}$ & 2.8$\times 10^{-11}$\\
$10^{3}$ & $0.1$ & $10^{-4}$ & 1 &   0.41 & 0.80  & 2.3$\times 10^{-12}$ & 1.3$\times 10^{-11}$ \\ \hline
$10^{3}$ & $0.1$ & $10^{-2}$ & 10 &  0.32 & 1.26  & 1.8$\times 10^{-12}$ & 2.5$\times 10^{-11}$\\
$10^{3}$ & $0.1$ & $10^{-2}$ & 3 &   0.22 & 0.82  & 2.5$\times 10^{-12}$ & 3.9$\times 10^{-11}$\\
$10^{3}$ & $0.1$ & $10^{-2}$ & 1 &   0.17 & 0.62  & 3.4$\times 10^{-12}$ & 5.9$\times 10^{-11}$\\
$10^{3}$ & $0.1$ & $10^{-2}$ & 0.3 &  0.16 & 0.65 & 5.4$\times 10^{-12}$ & 9.5$\times 10^{-11}$ \\ \hline
\end{tabular}
\end{center}
\end{table*}

Ambipolar diffusion can suppress MRI in low density regions of protoplanetary disks 
\citep[e.g.,][]{Blaes1994Local-shear-ins,Hawley1998Nonlinear-Evolu,Kunz2004Ambipolar-diffu,Desch2004Linear-Analysis,Bai2011Effect-of-Ambip,Simon2013Turbulence-in-t,Simon2013aTurbulence-in-t}.
If MRI is effectively suppressed in the e-heating zone, electric fields may not sufficiently grow to cause  electron heating.
The effectiveness of ambipolar diffusion is characterized by the ambipolar Elsasser number $\mathrm{Am} = \gamma_{i}\rho_{i}/\Omega$ \citep[e.g.,][]{Blaes1994Local-shear-ins,Lesur2014Thanatology-in-}, where $\gamma_{i}= \avg{\sigma_{in}v_{in}} / (m_{n}+m_{i})$ and $\rho_{i} = m_{i} n_{i}$.
According to MHD simulations including ambipolar diffusion, 
MRI-driven turbulence behaves as in the ideal MHD limit if $\mathrm{Am} \gg 1$, 
while ambipolar diffusion suppresses turbulence if $\mathrm{Am} \ll 1$ \citep[e.g.,][]{Bai2011Effect-of-Ambip}. 
Table \ref{tbl:Am} lists the values of ${\rm Am}$ as well as the ion abundance $x_i =n_i/n_n$ at the inner and outer edges of the e-heating zone before electron heating sets in ($E = 0$).
We find that $\mathrm{Am} \approx 0.2$--$0.7$, 
implying that ambipolar diffusion would moderately affect MRI turbulence in the e-heating zone. 
Therefore, MHD simulations including both electron heating and ambipolar diffusion 
are needed to assess which effect determines the saturation amplitude of MRI turbulence 
in these outer regions of the disks.

The Hall effect is also important at $r\sim 10$--$50 \, \mathrm{AU} $ \citep[see Figure 1 of][]{Turner2014Transport-and-A}.
The Hall effect can either damp or amplify magnetic fields, which depends 
on the relative orientation between the disk's magnetic field and rotation axis and
on the sign of the Hall conductivity \citep[e.g.,][]{Bai2014Hall-effect-Con,Wardle2012Hall-diffusion-}.
At relatively high gas densities ($n_{n} \ga 10^{10}~{\rm cm^{-3}}$), 
the Hall conductivity is usually positive \citep{Wardle1999The-conductivit,Nakano2002Mechanism-of-Ma,Salmeron2003Magnetorotation}, 
but can become negative when the number density of electrons is significantly lower than that of ions. 
Interestingly, our preliminary investigation shows that the Hall conductivity can indeed become negative as the electron number density is decreased by electron heating (Okuzumi et al., in prep.).
This suggests that electron heating might reverse the role of the Hall term. Whether this occurs under conditions relevant to protoplanetary disks will be studied in future work.

\section{Summary}\label{sec:Summary}
We have investigated where in protoplanetary disks the electron
heating by MRI-induced electric fields
affects MRI turbulence.
Our previous study \citepalias{Okuzumi2015The-Nonlinear-O} showed that
electron heating causes
a reduction of the electron abundance, and hence an amplification of
Ohmic dissipation,
when the recombination of plasma mainly takes place on dust grains
rather than in the gas phase.
To study where in disks this effect becomes important,
we constructed a simplified ionization model that takes into account
both recombination on dust grains and electron heating. The presented
model is computationally much less expensive than the original
electron heating model by \citetalias{Okuzumi2015The-Nonlinear-O} and
allows us to study the effects of electron heating for a wide range of
model parameters.
We then searched for locations in a disk where the enhanced Ohmic
diffusivity limits the saturation level
of MRI turbulence, which we call the ``e-heating zone,'' by using
analytic criteria for MRI growth.
Our results can be summarized as follows:

\begin{enumerate}

\item
We find that the e-heating zone can cover a large part of a
protoplanetary disk when tiny dust grains are abundant. 
For instance, in a minimum-mass solar nebula with 1\% of its mass consisting of
0.1-$\mu$m-sized dust grains, the e-heating zone extends out to 80 AU from
the central star (Figure \ref{fig:disk}; Section \ref{ssec:fid}). In
this case, MRI turbulence can develop without being affected by electron
heating only in the outermost region of $ r \gtrsim 80~{\rm AU}$.

\item
In the e-heating zone, the saturation level of MRI turbulence is
expected to be considerably
lower than that in fully MRI-active zones because the electron
heating sets an upper limit to the electric current density attainable
in MRI turbulence.
Our simple estimate based on scaling arguments (Section
\ref{sec:Suppress}) predicts that
for our fiducial disk model, the turbulence $\alpha$ parameter for MRI
turbulence
should be reduced to $\sim 10^{-5}$ and $10^{-3}$
at the inner and outer edges of the e-heating zone, respectively
(Figure \ref{fig:alp}).
This implies that the MRI is ``virtually dead'' deep inside the e-heating zone.

\item
Dust grains in the e-heating zone acquire a high negative charge due
to the frequent collisions with electrically heated electrons. This
strengthen the charge barrier against the growth of micron-sized
grains originally predicted by \citet{Okuzumi2009Electric-Chargi} (Figure \ref{fig:eel}; Section \ref{sec:CB}).
At midplane 35 AU in the fiducial model, the electric repulsion energy is larger than the collisional energy when the grain size is in the range of $\sim 0.08$--$0.5~{\rm \mu m}$.
We find that electron heating significantly enhances the charge barrier for submicron-sized grains.

\end{enumerate}

Our estimate of the turbulence strength in the e-heating zone (Equation \eqref{eq:alpha-2}) largely relies on
the scaling relations between turbulent quantities observed in
previous MHD simulations.
Although these scalings well predict the saturation level of MRI
turbulence without electron heating,
it is unclear whether they are still valid even in the presence of
electron heating.
Our future work will address this issue by performing MHD simulations
including electron heating.
We have also neglected the effect of magnetic fields on the kinetics of plasma,
which means that non-Ohmic effects such as the Hall effect
and ambipolar diffusion are excluded from our analysis.
However, these effects generally overwhelm Ohmic diffusion in outer
parts of protoplanetary disks. Our estimate indicates that 
ambipolar diffusion would moderately suppress MRI in the e-heating
zone (${\rm Am} \approx 0.2$--$0.7$; Section~\ref{ssec:Ha+Am}). Therefore, MHD
simulations including both electron heating and non-Ohmic
resistivities will be needed to assess which effect determines the
saturation amplitude of MRI turbulence in outer regions of the disks.
We will address these open questions step by step in future work.

\acknowledgments
The authors are grateful to Shigeru Ida, Taishi Nakamoto, and Shu-ichiro Inutsuka for the fruitful discussions and continuing support.
The authors also thank the anonymous referee for comments that improved the paper.
This work was supported by Grants-in-Aid for Scientific Research (\#23103005, 26400224, and 15H02065) from MEXT of Japan.

\end{document}